\documentclass{emulateapj}
\usepackage[usenames,dvips]{color}
\usepackage{amssymb, amsmath, amsbsy, epsfig, epsf}
\usepackage{longtable}
\usepackage{threeparttable}
\usepackage[figuresright]{rotating}

\def\cjaa{ChJA\&A}

\def\eg{{\em e.g.\ }}

\def\cm{{\rm\thinspace cm}}

\def\keV{{\rm\thinspace keV\thinspace}}

\def\pc{{\rm\thinspace pc}}

\def\chisq{\hbox{$\chi^2$}}

\def\pcmsq{\hbox{$\cm^{-2}\,$}}



\shorttitle{X-ray analysis of Was~61}
\shortauthors{Dou et al.}

\begin{document}

\title{X-ray spectral and temporal analysis of Narrow Line Seyfert 1 galaxy Was~61}

\author{Liming Dou\altaffilmark{1},
Ting-Gui Wang\altaffilmark{1},
Yanli Ai\altaffilmark{2},Weimin Yuan\altaffilmark{3}, Hongyan Zhou\altaffilmark{4,1}
 and Xiao-Bo Dong\altaffilmark{5,6,1} }

\altaffiltext{1}{Key Laboratory for Research in Galaxies and Cosmology,
Department of Astronomy, The University of Sciences and Technology of China,
Hefei, Anhui 230026, China}
\altaffiltext{2}{School of Physics and Astronomy, Sun Yat-Sen University, Guangzhou 510275, China }
\altaffiltext{3}{National Astronomical Observatories, Space Science Division, Chinese Academy of Sciences, Beijing 100012, China}
\altaffiltext{4}{Polar Research Institute of China, Jinqiao Road 451, Shanghai 200136, China}
\altaffiltext{5}{Yunnan Astronomical Observatories, Chinese Academy of Sciences, Kunming, Yunnan 650011}
\altaffiltext{6}{Key Laboratory for the Structure and Evolution of Celestial Objects, Chinese Academy of Sciences,
Kunming, Yunnan 650011, China}
\email{doulm@mail.ustc.edu.cn, twang@ustc.edu.cn}

\begin{abstract}

We present an analysis of spectrum and variability of the 
bright reddened narrow line Seyfert 1 galaxy Was~61 using 90 ks 
archival {\it XMM-Newton} data. The X-ray spectrum in 0.2-10\keV can be 
characterized by an absorbed power-law plus soft excess and an 
Fe K$\alpha$ emission line. The power-law spectral index remains 
constant during the flux variation. The absorbing material is 
mildly ionized, with a column density of 3.2$\times$10$^{21}$ cm$^{-2}$, 
and does not appear to vary during the period of the X-ray observation. If the 
same material causes the optical reddening (E(B-V)$\simeq$0.6 mag), 
it must be located outside the narrow line region with a dust-to-gas ratio similar to the average Galactic 
value. We detect significant variations of the Fe K$\alpha$ line during 
the observational period. A broad Fe K$\alpha$ line at $\simeq$6.7 keV with a 
width of $\sim$0.6\keV is detected in the low flux segment of 
the first 40 ks exposure, and is absent in the spectra of 
other segments; a narrow Fe K$\alpha$ emission line $\sim$6.4\keV 
with a width of $\sim$0.1 keV is observed in the subsequent 20 ks 
segment, which has a count rate of 35\% higher and is in the next day. 
We believe this is due to 
the change in geometry and kinematics of the X-ray emitting corona.
The temperature and flux of soft X-ray excess appear to correlate 
with the flux of the hard power-law component.
Comptonization of disc photons by a warm and optically thick inner disk
is preferred to interpret the soft excess, rather than the ionized reflection.

\end{abstract}

\keywords{galaxies: active - galaxies: Seyfert - X-rays: galaxies -
galaxies: individual (Was~61)}

\section{Introduction}
\label{sec:intro}

Strong X-ray emission is a general characteristic of active galactic nuclei 
(AGNs). Short timescale variability indicates that X-rays originate in a very compact region (i.e.,
a few gravitational radii in size; \eg, Mushotzky et~al.~1993; 
Fabian 2006, 2008; McHardy et~al.~2006). The observed X-ray spectrum of Seyferts consists of 
several components: a power-law component extending to at least a hundred keV, 
a soft X-ray excess component, a warm absorption component in roughly half 
the Seyfert galaxy population (e.g., Blustin et al.\ 2005), and a reflection component plus broad
 and narrow Fe K$\alpha$ lines and other weak emission lines. The power-law 
 component is thought to be produced by a hot corona located above the inner accretion disk, 
which is heated dynamically via a magnetic reconnection process or advection (Haardt \& Maraschi 1991; 
Zdziarski et al. 1994; Fabian et al. 2000). The soft X-ray excess component is usually thought to be 
a primary component with a debatable origin (e.g., Vasudevan et al. 2014 and references therein). These 
primary emissions are then modified by the gas surrounding the emission region and 
 further out, producing rich spectroscopic features. Incident X-rays on the thick 
 accretion disk or other materials will be reflected, and as a result, imprint on the X-ray spectrum 
 with a broad hump peaked at 30-40 keV (\eg, Pounds et al. 1990), as well as prominent absorption 
 edges and emission lines at lower energies, blurred by the Doppler broadening
and a gravitational redshift effect (\eg, Ross \& Fabian 1993; 
Gierli{\'n}ski \& Done 2004; Crummy et al. 2006).
 Depending on the proximity of the gas to the central black hole (BH),
emission lines can be broad or narrow. Gas in the line of sight, if present, will 
produce either ionized or neutral absorption lines and edges. These features 
have all been observed in the X-ray spectra of AGNs and can be temporally variable 
(\eg, Mushotzky et~al.~1993; Petrucci et al. 2002; Markowitz, Edelson, 
\& Vaughan 2003; Turner et al. 2008; Patrick et al. 2012).

The relatively high abundance and high fluorescent yield make the Fe K$\alpha$ 
the strongest emission line in the X-ray reflection light in the hard X-ray band. 
The observed Fe K$\alpha$ usually consists of a narrow component (FWHM$<$10$^4$
km~s$^{-1}$) and a broad component (FWHM$>$10$^4$ km~s$^{-1}$, e.g., Shu et al. 2010). 
The first unambiguous broad Fe K$\alpha$ line was discovered by {\it ASCA} in MGC-06-30-15 
(Tanaka~et~al.~1995), and the average fraction of AGNs with broad Fe K$\alpha$ lines 
is about 50$\%$ or more (Porquet et~al.~2004; Guainazzi et~al.~2006; Nandra et~al.~2007; 
de~La~Calle~P{\'e}rez~et~al.~2010; Patrick et al. 2012; 
Walton et al. 2013 and references therein).
The observed red-skewed broad line can be naturally
interpreted as an emission line from a relativistic accretion disk around a 
BH, where its unique profile is shaped by the Doppler broadening, relativistic 
beaming, and gravitational redshift effect, although other models cannot be 
excluded completely, based on the X-ray spectrum alone (\eg, 
Nandra~et~al.~2007; Turner~et~al.~2007, 2008; Turner~\&~Miller~2009). 
Because the innermost radius of an accretion disk is closely related to
the BH spin, within such a framework the broad Fe K$\alpha$ line profiles have 
been used to measure the BH spin and disk inclination (see the reviews in 
Fabian~et~al.~2000; Fabian 2006, 2008; and some recent references 
for measuring the BH spin in AGNs, \eg, Brenneman et al. 2011; Nardini et al. 2011;
Lohfink et al. 2012; Patrick et al. 2012; Walton et al. 2013 and Liu et al. 2015).
Narrow Fe K$\alpha$ lines, peaking at 6.4 keV or 6.7 keV, are commonly observed 
in local AGNs (Yaqoob~\&~Padmanabhan~2004; Shu~et~al.~2010; Shu~et~al.~2012 and 
references therein). They are thought to be formed further out, such as in the outermost 
region of the accretion disk, the broad line region, and/or torus
(Jovanovi{\'c}~2012). But the exact location of such gas is not yet clear.

Warm absorptions and soft X-ray excesses are two common features in the soft 
X-ray band. In high resolution X-ray spectra, rich narrow absorption lines 
and photoelectric absorption edges from partially ionized materials are often
detected. The absorption lines are usually blueshifted, which suggests that the 
ionized gas outflows from the galactic nucleus. At low resolutions, these 
absorption lines and edges blend into a broad deficiency at 0.5-1.5 keV (\eg,
Halpern~1984; Reynolds~1997; George~et~al.~1998). The temporal variability of 
absorption indicates that the warm absorbers locate within a few ten parsecs of 
the nucleus (\eg, Kaspi~et~al.~2000; Krolik \& Kriss 2001; Blustin~et~al.~2005). 
Intensive monitoring of a few bright Seyfert galaxies and subsequent 
photoionization modeling suggest that the absorbing gas is multi-phase 
with a range of ionization (\eg, Kaspi et al. 2002; Krongold et al. 2007; Arav et al. 2015).

The origin of the soft X-ray excess is not well understood either. The feature can 
be modeled as a black-body, but the temperatures derived from the spectral fits 
are in the range of 0.1$-$0.2 keV, which are far higher than the expected maximum 
temperature from the inner region of an optically thick accretion disk (\eg, 
Czerny~et~al.~2003; Gierli{\'n}ski~\&~Done~2004; Ai et~al.2011). Thus, thermal 
emission from an accretion disk is not favored except in a Narrow Line Seyfert 
1 (NLSy1; RX J1633+4718, Yuan et al. 2010) or two super soft X-ray sources 
(2XMM J123103.2+110648, Terashima et al.~2012; RX~1301.9+2747, Sun et al. 2013). 
Comptonization of thermal emission by a thick warm layer can 
increase the temperature to the observed range (Czerny~\&~Elvis~1987; Wandel~\&
Petrosian~1988; Shimura~\&~Takahara~1993; Czerny et al. 2003). On the other 
hand, the soft X-ray excess may not be a primary continuum component, insteand it may be 
due to the atomic features in the reflection component from an ionized
relativistic accretion disk or smeared absorption edges (Gierli{\'n}ski~\&~Done~2004; 
Ross~\&~Fabian~2005; Pal \& Dewangan 2013). 
Magnetic re-connection has also been proposed in a recent study (Zhong~\&~Wang 2013).

Thanks to the long-time X-ray observation by {\it XMM-Newton}, {\it Suzaka} and {\it NuSTAR}, 
the reverberation technique is broadly used to explore the geometry and physical 
properties of the emission region. Recently, reverberation time delays between 
the power-law component and soft excess have been measured in some Seyfert 
galaxies (\eg, Fabian et al. 2009, 2012; Zoghbi et al. 2010; 
de Marco et al.2011; Emmanoulopoulos et al. 2011; Zoghbi \& Fabian 2011; 
Cackett et al. 2013, 2014; Kara et al. 2013a). More recently, some authors have reported 
the detection of Fe K lags (\eg, Zoghbi et al. 2012, 2013; Kara et al. 2013a,
2013b, 2014; Marinucci et al. 2014). The reflection origin of the soft excess, which 
is similar to the broad Fe K emission, is supported in many individual Seyfert 
galaxies. In particular, a relativistic iron line was detected in the lag 
spectra of MCG-05-23-16 on three different timescales, allowing the emission
from different regions around the BH to be separated (Zoghbi et al. 2014). 
 Since the launch of {\it NuSTAR} on 2012 June 13, its broadband (3-79 keV) 
and high signal to noise ratio (S/N) spectra allow an accurate separation 
of the reflection and primary continua, as well 
as a more precise determination of the ionization state of the 
reflector (e.g., Risaliti et al. 2013; 
Brenneman et al. 2014a, 2014b; Marinucci et al. 2014).

However, the reverberation technique usually needs a long uninterrupted 
exposure on a high count rate source. Joint spectral features and temporal
analysis will provide a useful constraint on these processes. In this paper, 
we present a detailed analysis of the bright Seyfert galaxy, Was~61, to 
investigate the possible origins for both soft X-ray excess and Fe K$\alpha$ 
line emissions. Was~61 is a NLS1 at redshift 0.0435 (Grupe~et~al.~1999a; 
Grupe~et~al.~2004; Du~et~al.~2014). The {\it ROSAT} spectrum shows typical properties 
of a NLS1, (i.e., strong soft X-ray excess and strong variability; \eg, Grupe~et~al.~2001;
Cheng~et~al.~2002; Bian~\&~Zhao~2003). There is an indication for warm 
absorption along the line of sight (Grupe~et~al.~1999b). Strong and broad Fe K 
line emission at around 6.4\keV is presented in the {\it XMM-Newton} spectra
(Bianchi~et~al.~2009; Zhou~\&~Zhang 2010). Although Was~61 has the typical 
observed features in the X-ray spectra, there is no detailed study for this object.
In this paper, we present a detailed analysis of the two {\it XMM-Newton} 
observations of Was~61, specifically on the variation of Fe K emission, 
soft X-ray emission, and warm absorber.

The paper is organized as follows. In section 2, we describe the data reduction. 
The spectral modeling, light-curve analysis, and spectral variation are discussed in section 3 
and 4. The results are discussed in section 5 and concluded in sections 6.
Throughout this paper, luminosities are calculated assuming a ${\Lambda}CDM$ 
cosmology with $\Omega_{M}$ = 0.27, $\Omega_{\Lambda}$ = 0.73, and a Hubble 
constant of $H_{0}$ = 70 km s$^{-1}$ Mpc$^{-1}$, corresponding to a luminosity distance of 
D=192.5 Mpc to the galaxy. We adopt a Galactic absorption column density of 
$N_{H}$ = 1.35 $\times$ $10^{20}$ \pcmsq (Kalberla et~al.~2005) in the direction 
of Was~61, and a BH mass of $4.6^{+1.5}_{-1.2}\times$10$^{6}$ M$_{\sun}$
(Du et~al.~2014).

\section{Observations and Data Reduction}
\label{sec:data}

Was~61 was observed with all instruments onboard the {\it XMM-Newton} telescope in 
two consecutive orbits during  2005 June 20 (Observation ID 0202180201, 
hereafter obs1) and  2005 June 23 (Observation ID 0202180301, hereafter 
obs2) for about 80 ks and 12 ks, respectively. Both {\it pn} and {\it MOS} detectors 
were operated in the full-frame mode with a thin filter and both two reflection 
grating spectrometers ({\it RGS}s) were operated in spectroscopy mode. Due to the superior 
statistical quality of the {\it pn} camera (Str{\"u}der~et~al.~2001), the following 
analysis will rely mainly on data from the {\it pn} camera, 
although the {\it RGS} and {\it MOS} data are occasionally used, mainly for clarification purposes.

The data reduction was processed using the standard Science Analysis System 
(SAS, v12.0) with the most updated calibration files. The Observation Data Files 
are processed to create calibrated events files with ``bad'' (\eg, ``hot,'' ``dead,''
``flickering'') pixels removed. The time intervals of high flaring background 
contamination are identified and subsequently eliminated following the standard 
SAS procedures and thresholds. Source counts are extracted from a circular region 
of radius $40^{\prime \prime}$,  while the background counts are extracted from 
a source-free region with the same radius. We 
check the pipe-up effect using SAS task {\it epatplot}, and find that it affects 
the 0.5-2 keV band at about a 10\% level. Thus, in the following analysis, we 
will use single events (PATTERN = 0, FLAG = 0) only. This results in about 
271.1 and 35.5 thousand net source counts in 0.2-10 keV band for {\it pn}, and net 
exposure times of 54.8 and 9.0 ks for obs1 and obs2, respectively.
Light curves with a bin size of 200 sec are extracted for obs1 and obs2. 
We run the task {\it epiclccorr} to subtract the background and subsequently 
correct for instrumental effects on lightcurves. 

For the {\it RGS} data, we run the task {\it rgsproc} with default parameters 
to extract the calibrated first-order spectra and responses. 
We mask a few narrow bands that might be instrumental
absorption features (Pollack 2015). 
This results in total 23.4 thousand net source counts between 7 and 37.9\AA\
in two RGS spectra.

 During $obs1$, the Optical Monitor ({\it OM}) was observing in the UV with filters 
{\it UVW1}, {\it UVM2}, and {\it UVW2}, corresponding to the effective wavelengths of 2910, 2310, 
and 2120\AA, respectively. We use the task {\it omichain} to process the {\it OM} data. 
Ten exposures were taken, with four in {\it UVW1}, five in {\it UVM2} and one in {\it UVW2}. 
We do not find any significant variation in the UV band.

 Was~61 was observed with the {\it ROSAT PSPC-b} with an exposure time of 18.8 ks 
and an offset of $\sim42^{\prime}$ in 1991 December (ObsID: 600129). We retrieved the {\it ROSAT} 
data from the archive and use the {\it XSELECT} tool of {\it FTOOLS} for data analysis.
The X-ray spectrum is extracted from a circular region with a radius of $300^{\prime \prime}$, and a background 
spectrum was extracted from a source-free circular region with the same radius and offset. 
This results in $\sim$9.2 thousand net source counts.

 Since $obs1$ has a much longer cleaned exposure time than $obs2$ and {\it ROSAT}, 
we firstly focus the spectral analysis on $obs1$. $Obs2$ and {\it ROSAT} are only 
used in variability studies. The {\it pn} and {\it ROSAT} spectra 
are then grouped to ensure at least 25 counts per bin. 
The two {\it RGS} spectra are grouped to five channels or $\sim$0.05\AA\ per bin. 
This gives a total approximately 0.95 thousand bins with an average of 
$\sim$24.7 net counts each. 
The spectral fitting is performed using XSPEC (v.~12.8; Arnaud 1996), 
and applied to the $\chisq-statistics$ (Avni 1976). The neutral absorption 
column density is fixed at the Galactic value ($\it{tbnew}$ in XSPEC, Wilms~et~al.
~2000). The uncertainties are given at 90\% confidence levels for one 
interesting parameter.

\section{The X-ray Spectral Analysis}

\subsection{Hard X-Ray Spectrum and Fe K$\alpha$ Line} \label{hard}

Initially we fit the X-ray spectrum in the hard band (2-10\keV) using 
a power-law model with the Galactic absorption. The best fit is acceptable 
with the photon index $\Gamma=2.13\pm0.04$ (see Table \ref{tab:poga1}). 
Yet the systematic deviations around 6.0-8.0\keV are still visible in the 
residual spectrum. Adding a Gaussian line steepens the photon index to
2.17$\pm$0.04 and improves the fit significantly, with a chance probability 
of $4.3\times10^{-4}$ according to the $F-test$. The best-fit line center is 
6.43$\pm$0.25\keV in the source rest frame, thus associated with Fe K$\alpha$. 
It is a broad emission line with a width of $\sigma$=0.49$\pm$0.30\keV. 
Additional narrow Fe K$\alpha$ is not required according to the F-test with a $P_{null}$=87\%.
These results are consistent with those in the literature (\eg, Bianchi~et~al.
~2009; Zhou~\&~Zhang~2010).

\subsection{Soft X-Ray Excess and Warm Absorption} \label{full}

We extrapolate the above best-fit model (power-law plus the broad 
Fe K$\alpha$) to the full 0.2-10\keV band. Excesses below 0.7\keV and 
deficiencies around 0.7-1\keV are prominent. Adding a blackbody 
component to the model does not yield an acceptable 
fit with \chisq=1732 for 877 dof and evident absorption features in the
residuals. We then use a Gaussian line to characterize the absorption 
feature. The model ($\it{tbnew*gabs*(zbbody+powerlaw+zgauss)}$) gives 
an acceptable fit (\chisq=983.5 for 874 dof), in which the absorption line 
is centered at 0.79$\pm$0.01 keV with its width of 0.09$\pm$0.01 keV.
However, there are still systematic residuals in the soft bands. 
As a trial run, we use an absorption edge 
($\it{tbnew*zedge*(zbbody+powerlaw+zgauss)}$) 
to replace the absorption line. Although the fit is improved (\chisq
= 963.9 for 875 dof), the systematic residuals are still visible.

\begin{figure}[htbp]
\centering
\begin{minipage}{0.8\textwidth}
\centerline{\includegraphics[width=0.8\textwidth,angle=-90]{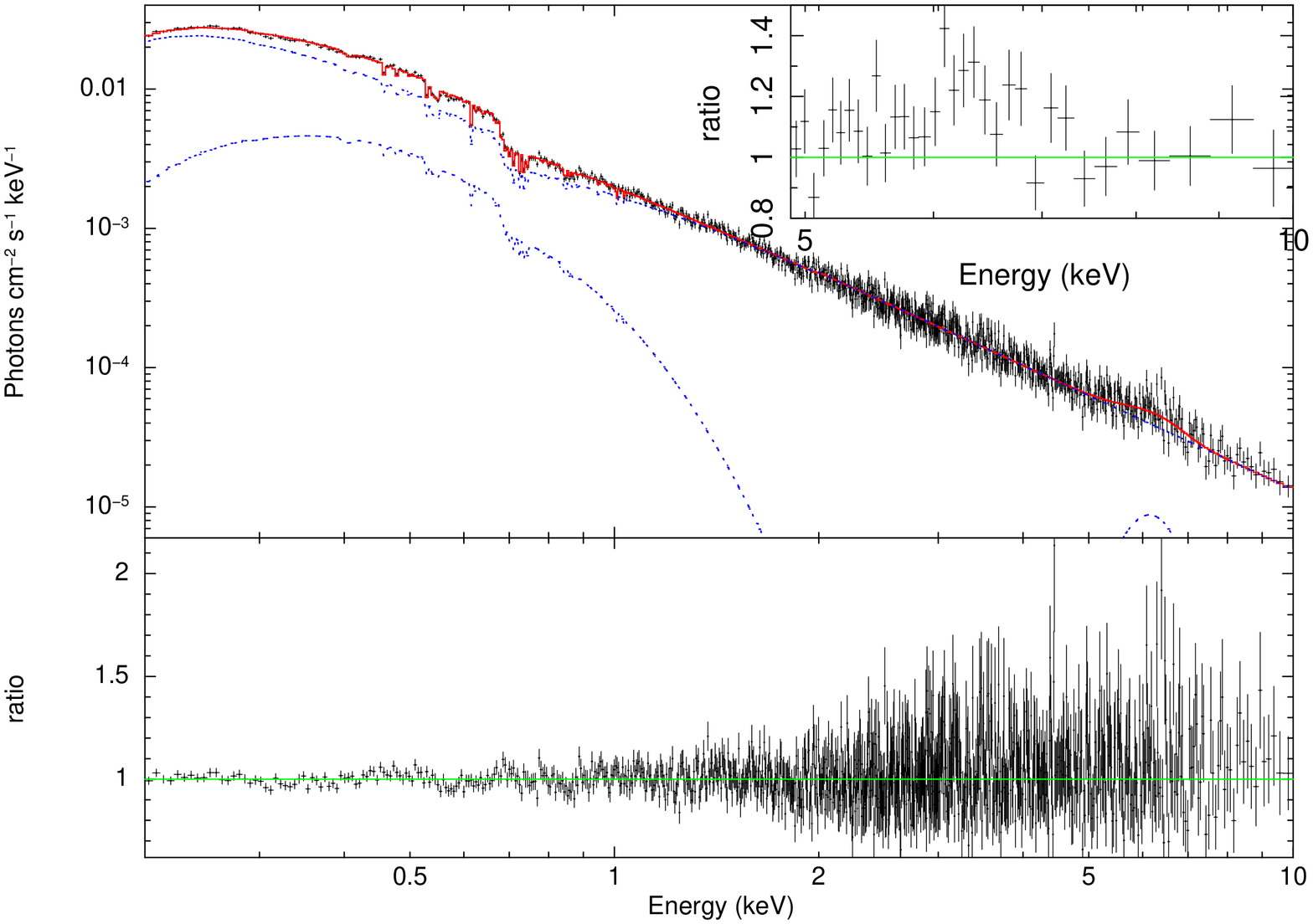}}
\end{minipage}
\caption{{\it XMM-Newton pn} spectrum of Was 61 with fitted model (red) and data to model ratio.
 The model is a power-law (blue dot) plus a broad Gaussian component (blue dot) 
and a blackbody component (blue dot), which includes the Galactic absorption 
and a warm absorption. The top right is a zoomed plot in 5-10\keV for the 
data to model ratio with a power-law model. }
\label{fig:spectr}
\end{figure}

The energy of the absorption line or edge suggests that the absorbing gas 
is ionized, which leads us to use a physical warm absorption model. We adopt 
$\it{warmabs}$, which is an analytic model obtained by fitting $XSTAR$ 
(Kallman \& Bautista 2001) results. The pre-calculated level populations 
were generated by $XSTAR$ (version 2.2), assuming a gas density of 
10$^4$~cm$^{-3}$ and a power-law ionizing continuum with $\Gamma$ = 2.2, 
which matches approximately the observed power-law component.  We fix the 
turbulent velocity at 30 km s$^{-1}$ because it cannot be resolved by 
{\it pn} or {\it RGSs} and the final result is not sensitive to its exact value. 
 We begin to model the warm absorber with a zero velocity
offset relative to the systematic one.
The fit ($\it{tbnew*warmabs*(zbbody+powerlaw+zgauss)}$) 
is acceptable (Figure~\ref{fig:spectr} and Table~\ref{tab:bestmodel}).

\begin{figure}[htbp]
\centering
\begin{minipage}{0.43\textwidth}
\centerline{\includegraphics[width=0.8\textwidth,angle=-90]{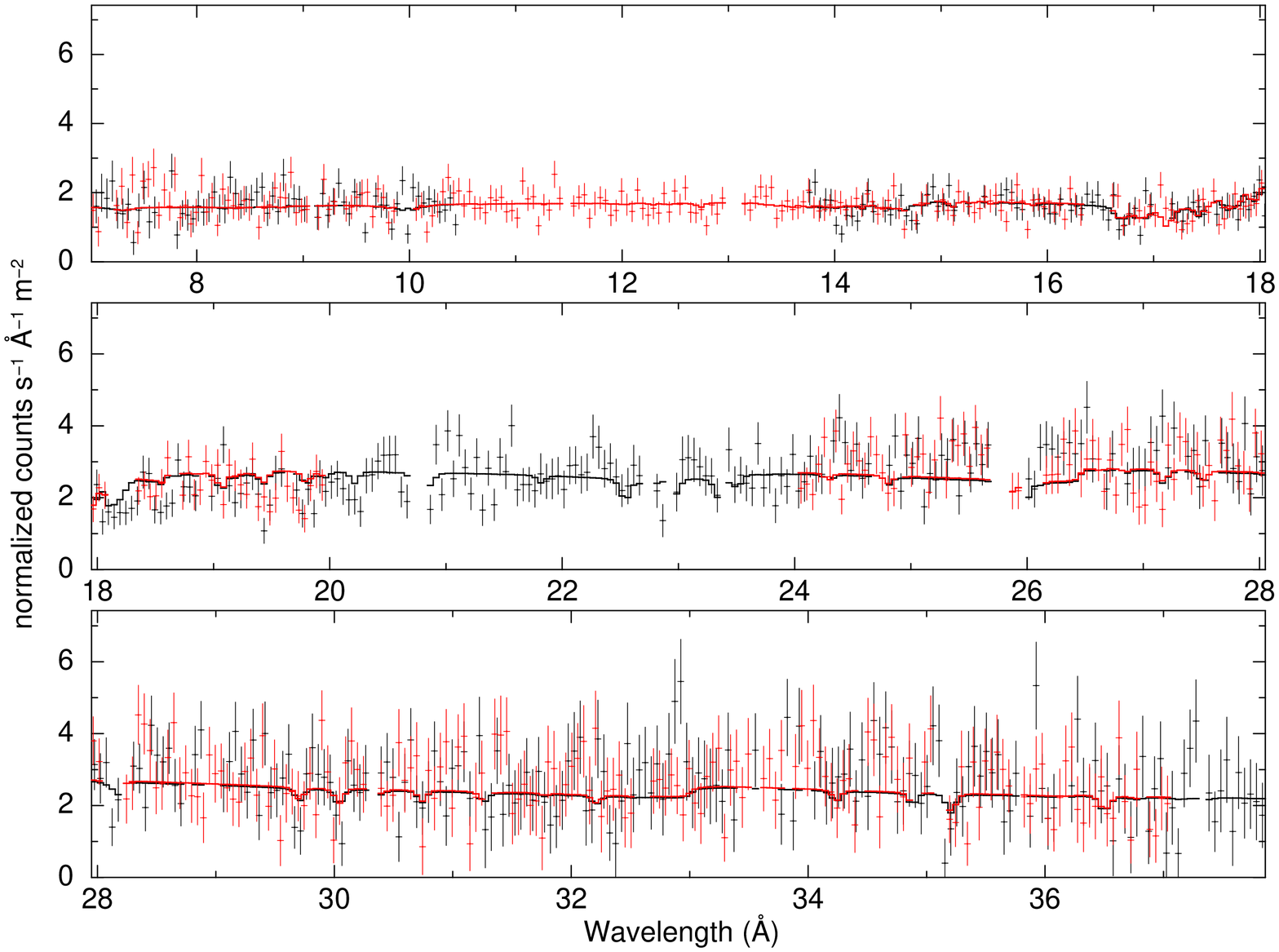}}
\end{minipage}
\hfill
\begin{minipage}{0.43\textwidth}
\centerline{\includegraphics[width=0.8\textwidth,angle=-90]{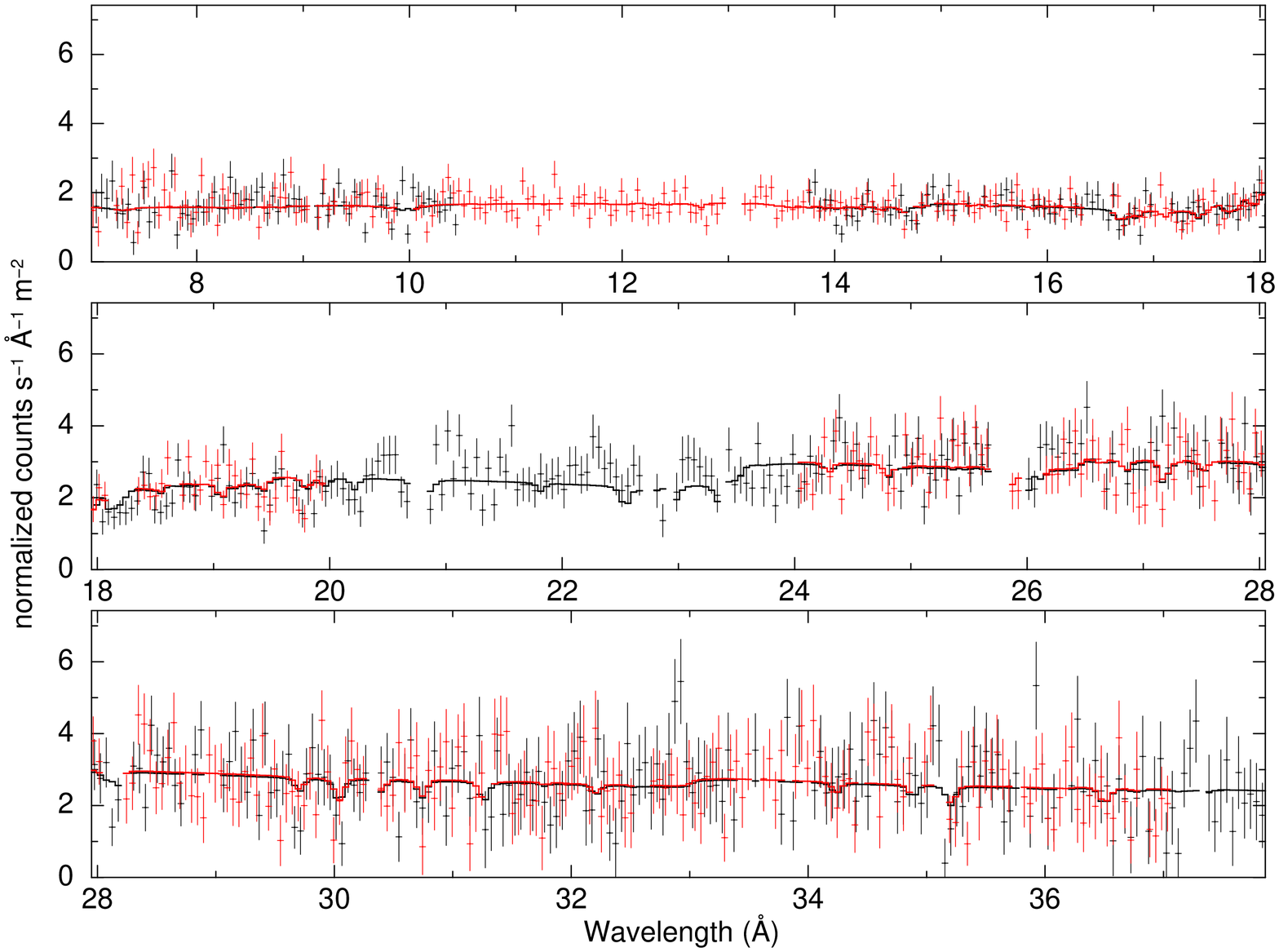}}
\end{minipage}
\caption{The {\it XMM-Newton RGS} 1 (black) and {\it RGS} 2 (red) spectrum. Left panel:
the same model as used in Figure 1;
Right panel: the same model as used in the left one, but with its C, N, O, and 
Fe abundances of the absorber allowed to vary freely.
}
\label{fig:rgs}
\end{figure}

Because RGS has much better energy resolution, we can examine the absorption 
and emission features in detail to try to further constrain the properties 
of the absorber. We initially fit {\it RGS} 1 and 2 spectra (Figure~
\ref{fig:rgs}), simultaneously, using the same model but without the broad 
Fe K$\alpha$ line, which is out of the {\it RGS} spectral coverage. We fix 
$\Gamma$ and {\it{kT}} of the blackbody to the best-fit values for 
{\it pn} data because they were much better constrained there. This 
results in a total $\chi^2/dof$ of 1142.6/951 (Table~\ref{tab:bestmodel}). 
We spot several absorption lines in the spectra, which are also evident in 
the current model at 35.16\AA, 30.80\AA, 30.02\AA, 29.67\AA, 19.81\AA, 
19.43\AA\, and 18.54\AA\ in the observed-frame, corresponding to C {\sc vi} 
$Ly-\alpha$ at 33.69\AA, 
N {\sc vi} He-$\alpha$ (f) at 29.52\AA, N {\sc vi} He-$\alpha$ (r) at 28.77\AA, 
C {\sc vi} Ly-$\beta$ at 28.44\AA, O {\sc viii} $Ly-\alpha$ at 18.99\AA, 
O {\sc vii} He-$\beta$ at 18.62\AA\, and O {\sc vii} He-$\gamma$ at 17.76\AA\ 
in the rest frame, respectively. However, some of the observed lines appear 
stronger than in the model, especially the O {\sc viii} Ly-$\alpha$,  
O {\sc vii} He-$\beta$ and O {\sc vii} He-$\gamma$ lines.
To assess the significance of the excess of absorption lines, we fix the 
parameters of the current warm absorber, blackbody, and power-law components,
and add three extra-absorption lines, 
relying on O {\sc viii} $Ly-\alpha$, O {\sc vii} He-$\beta$ 
and O {\sc vii} He-$\gamma$ lines. 
We fix the line centers to their rest frame wavelength and width to 0.05\AA\, 
the bin size. The fit gives the line depths of O {\sc viii} $Ly-\alpha$, 
O {\sc vii} He-$\beta$ and O {\sc vii} He-$\gamma$ 
at values of $0.0033^{+0.0013}_{-0.0016}$, $0.0035^{+0.0013}_{-0.0016}$, 
and $0.0026_{-0.0012}^{+0.0015}$, with $\Delta\chi^2$ values of 26.5, 26.8, and 15.9, 
respectively. We then fix the line depths at the best-fit values, but allow the line centers to vary freely.
The fit gives the line centers at values of $18.94\pm0.04\AA$, $18.63\pm0.03\AA$, and $17.76\pm0.05\AA$, respectively.
These lines are detected at a high significance level with a total 
$\Delta\chi^2=69.2$ for $\Delta\nu=3$ ($P_{null}=2\times10^{-13}$). 
If the redshift is allowed to vary freely, the fit does not significantly 
improve with an $\Delta\chi^2\sim1.2$ for $\Delta\nu=1$. Thus, the redshift of the 
warm absorber is consistent with the systematic one.  

The excessive optical depths of these lines can be caused by large abundances 
of relevant elements, the presence of an additional component of higher ionization 
absorber, or a combination of the two. To test these ideas, we fix the column 
density and ionization parameter of the current (or first) warm absorber 
and the normalization of the current blackbody and power-law components, 
in order to keep the continuum absorption consistent with the $pn$ model, 
and fit the RGS spectra with following two schemes: (1) allow C, N, and O 
abundances of the absorber to vary freely; or (2) adding another highly ionized 
absorber component.   
In both cases, the fit is significantly improved with respect to the best 
{\it pn} model. The best fit for the free abundance model converges to the abundances 
of C, N, and O at $0.29\pm0.14$, $3.64\pm0.50$, and $0.89\pm0.07$, 
respectively, with 
$\Delta\chi^2=90.6$ for $\Delta\nu=3$ ($P_{null}=7\times10^{-17}$). 
The second scheme results in a column density N$_H$$\sim$$10^{24}$ cm$^{-2}$ 
and $\xi$$\sim$10$^{4}$ erg cm s$^{-1}$ for the second warm 
absorber component, with a $\Delta\chi^2=17.9$ for $\Delta\nu=2$. It is 
worse than scheme 1 at $P_{null}=2\times10^{-15}$, according to the $F-test$.
If dust depletion is important, then some other elements will also be depleted, 
such as Fe. We then allow the Fe abundance to vary freely. It results in the 
abundances of C, N, O, and Fe at $0.63_{-0.20}^{+0.15}$, $3.59\pm0.51$,  
$0.97\pm0.08$, and $0.02_{-0.02}^{+0.42}$, respectively, with $\Delta\chi^2=12.3$ 
for $\Delta\nu=1$ ($P_{null}=8\times10^{-4}$). Thus, the Fe abundance is also 
consistent with the dust depletion.

The low S/N ratio does not allow us to explore other weak lines. 
The column density and ionization parameter $\xi$ of the warm absorber 
derived from the {\it RGS} spectra are slightly higher than the ones 
derived from the {\it pn} spectrum. Considering potential calibration 
uncertainty in the continuum slope between the two instruments, 
these values may be considered as being consistent. In the further fitting 
for {\it pn} and {\it ROSAT} spectrum, we still use the one warm absorber 
model with solar-type abundance to simplify to model the absorption of Was 61. 

This model without the broad Fe K$\alpha$ line is also used to fit the 
{\it ROSAT} spectrum. We fix the $\Gamma$ to 2.2, but allow {\it{kT}}, 
the column density, and ionization parameter of warm absorber to vary freely. 
The best fit is acceptable and converges to a lower blackbody temperature 
and a lower ionization parameter of the warm absorber (see Table~
\ref{tab:bestmodel}), while the absorbing column density is constant within 
its uncertainty. Note that 0.2-2 keV flux (4.1$\times$10$^{-12}$ erg cm$^{-2}$~s$^{-1}$) 
during the {\it ROSAT} observation is about half the {\it pn} 
flux (8.1$\times$10$^{-12}$ erg cm$^{-2}$~s$^{-1}$) in the same 
band, so the change in the ionization parameter may be entirely driven 
by the variations in the ionizing continuum flux.

\section{Temporal Analysis and Spectral Variability}

\subsection{X-Ray Variability}

\begin{figure}
\begin{minipage}{0.33\textwidth}
\centerline{\includegraphics[width=0.9\textwidth]{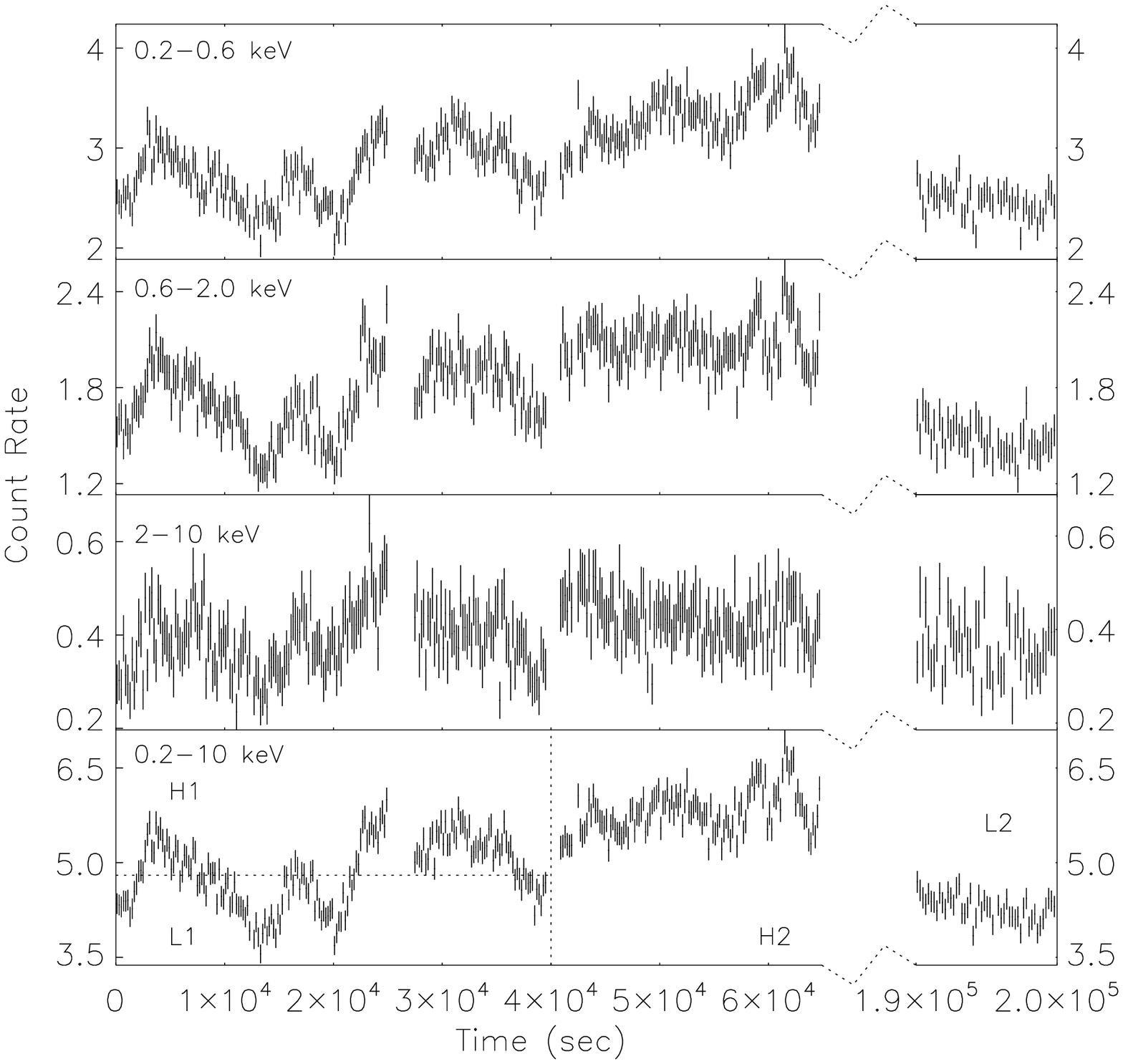}}
\end{minipage}
\hfill
\begin{minipage}{0.3\textwidth}
\centerline{\includegraphics[width=0.9\textwidth]{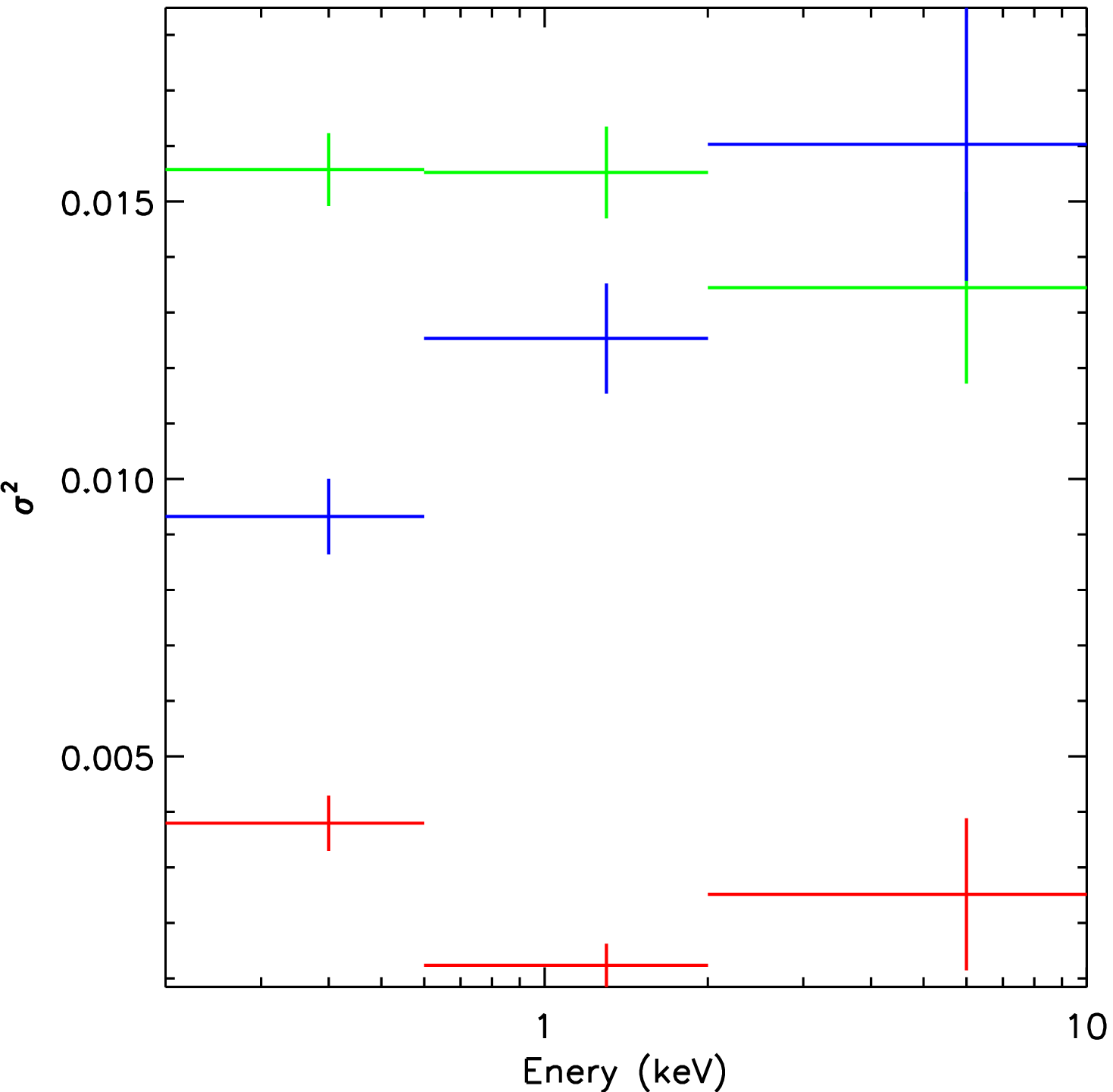}}
\end{minipage}
\hfill
\begin{minipage}{0.3\textwidth}
\centerline{\includegraphics[width=0.9\textwidth]{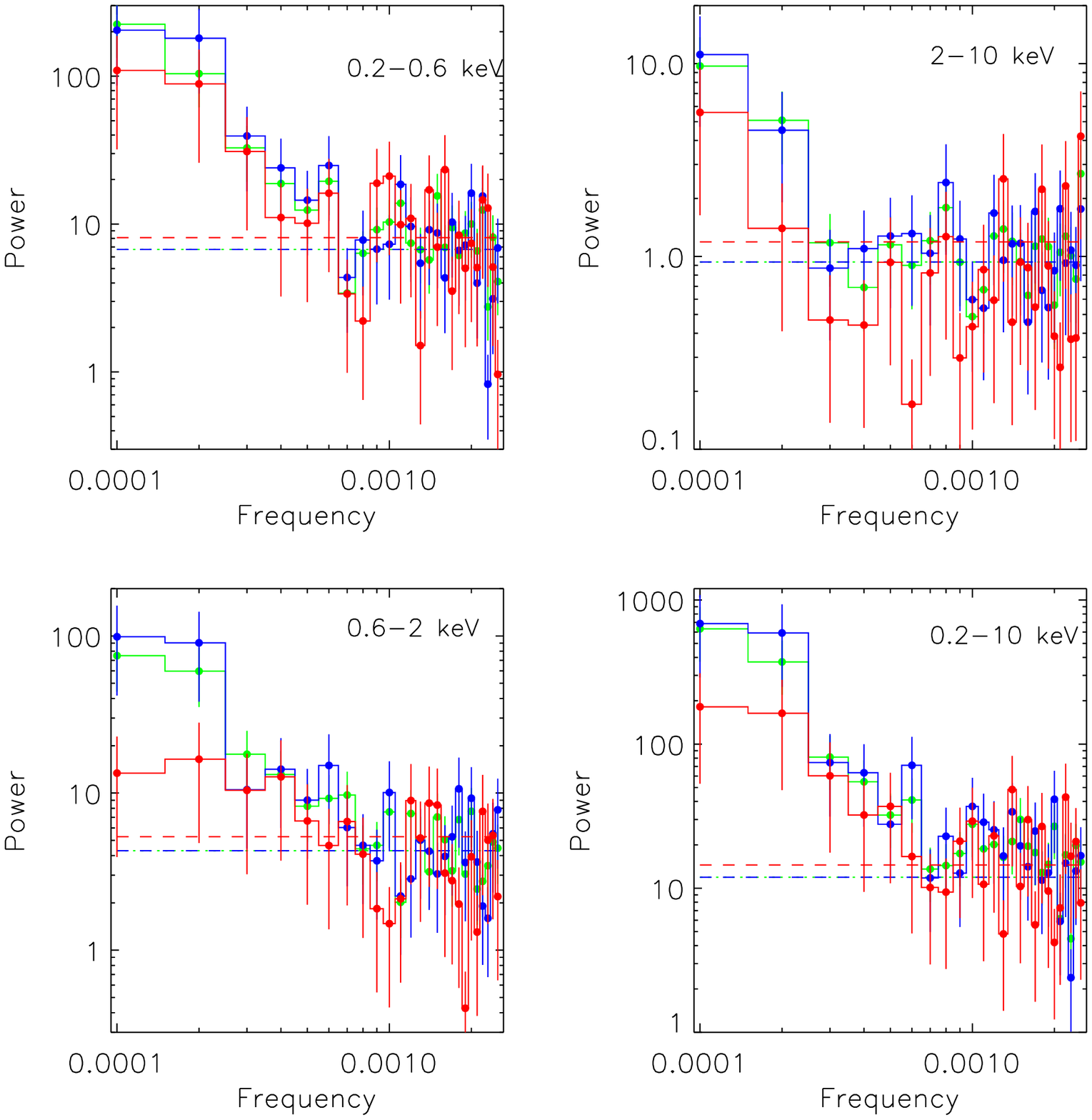}}
\end{minipage}
\caption{Left panel: {\it XMM-Newton} lightcurves for Was~61 in each band.
The time bin size is 200 s. The left is Obs1, the right is Obs2.
The division of the four segments flagged as $L1$, $H1$, $H2$,
and $L2$ state is illustrated in the 0.2-10\keV light curve.
Middle panel: normalized excess variance of the Obs1 observation as a function of energy 
(the light curves is rebinned to 1 ks before calculating the NXS).
Right panel: power spectral densities of Obs1 in each energy band, 
the lightcurve is rebinned to 200 s before creating the power spectral densities, 
the dashed line is the Poisson noise level (which is given as Vaughan et al.\ 2003). 
Green symbols are from the full exposure of Obs1, blue symbols are from the first 
40 ks segment ($L1$+$H1$); and red symbols are from the last 20 ks segment ($H2$).
\label{fig:lv}}
\end{figure}

The lightcurves in the 200 s bin in 0.2-0.6, 0.6-2, 2-10, and 0.2-10 keV 
bands during obs1 and obs2 are shown in the left of Figure~\ref{fig:lv}. It is 
evident that Was~61 varies on timescales of a few ks in all of these bands, 
which is similar to most NLS1s. Normalized excess variances (NXS, Ponti et~al.~2012) 
are on the same level (i.e., 13\%) for all bands (middle panel of Figure~\ref{fig:lv}) 
in obs1. When splitting obs1 into two segments, there seems to be a trend that NXS 
increases from soft to hard X-ray bands during the first 40 ks exposure of 
obs1, but $\chi^{2}-test$ suggests that the difference is not statistically 
significant ($P_{null}=22\%$).
The NXS of the first 40 ks is a factor of 2-5 higher than that of the 
last 20 ks exposure of obs1, 
depending on the energy band. Since the lengths of
the two segments are different, the large NXS of the first segment may be attributed 
to additional power at lower frequencies. Thus, we calculate the power spectral 
densities (PSDs) for the entire, first 40 ks, and last 20 ks lightcurves of $obs1$ 
(right panel of Figure~\ref{fig:lv}) using the IDL code $``dynamic\_psd.pro''$ 
(written by Vaughan, see Vaughan~et~al.~2003). Significant power is detected between 
(1-5) $\times$ 10$^{-4}$ Hz in the total, 0.2-0.6 keV, and 0.6-2 keV bands. Due to much 
worse statistics, only PSD in the lowest bin is significantly above zero in the 
2-10 keV band. At lower frequencies (1-2 $\times$ 10$^{-4}$ Hz), the PSD of the 
last 20 ks is significantly smaller than that of the first 40 ks in 0.6-2 keV 
band (at 90\% confidence level). There is also no significant difference 
either at higher frequencies or in the 0.2-0.6\keV band, although the peak power 
in that band is slightly higher than in the 0.6-2.0\keV band, 
suggesting that soft X-ray variability is more stationary.

\subsection{Temporal Spectral Analysis}\label{sec:tempsp}

To find what causes the energy dependence of variability amplitudes, we divide 
the first 40 ks exposure of obs1 into a low ($L1$) and a high flux ($H1$) 
states according to the count rate, and examine the variability of different 
spectral components. The division at a count rate of 4.8 cts~s$^{-1}$ is
chosen to guarrantee that each state has roughly the same exposure time (20 ks). For 
convenience, we flag the last 20 ks of $obs1$ as $H2$ and $obs2$ as $L2$ 
(marked in the left of Figure~\ref{fig:lv}). The mean count rates of these 
segments in 0.2-10\keV are 4.11$\pm$0.02, 4.96$\pm$0.02, 5.49$\pm$0.01, 
and 3.92$\pm$0.02 cts s$^{-1}$, respectively. 
The spectrum in each segment is then extracted from the clean event file. We 
also extract the spectrum of entire obs1 for comparison. The effective area 
and response files are generated using the most recent calibration files 
(updated to 2013 January).

\begin{figure}[htbp]
\begin{minipage}{0.43\textwidth}
\centerline{\includegraphics[width=0.8\textwidth,angle=-90]{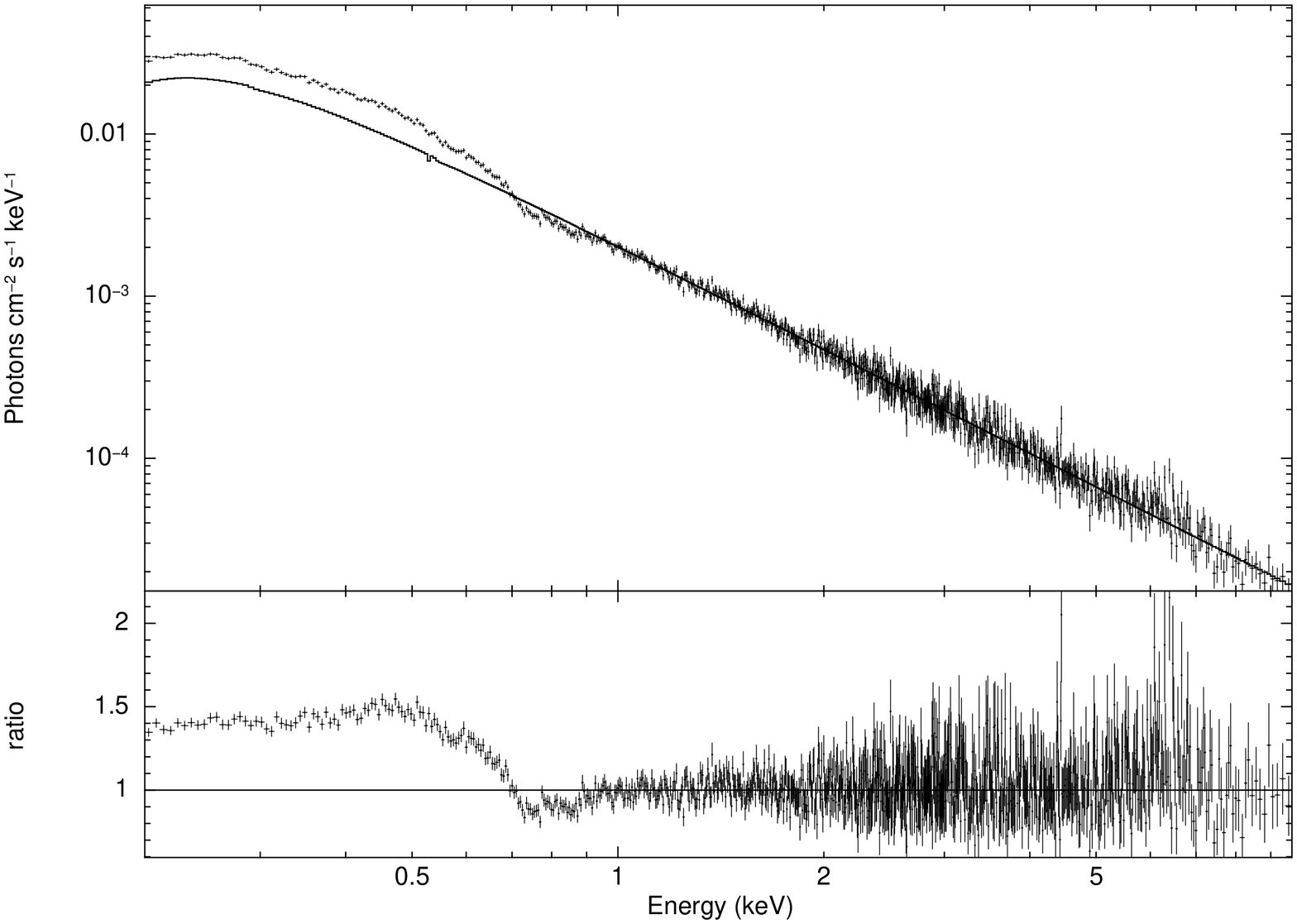}}
\end{minipage}
\hfill
\begin{minipage}{0.43\textwidth}
\centerline{\includegraphics[width=0.8\textwidth,angle=-90]{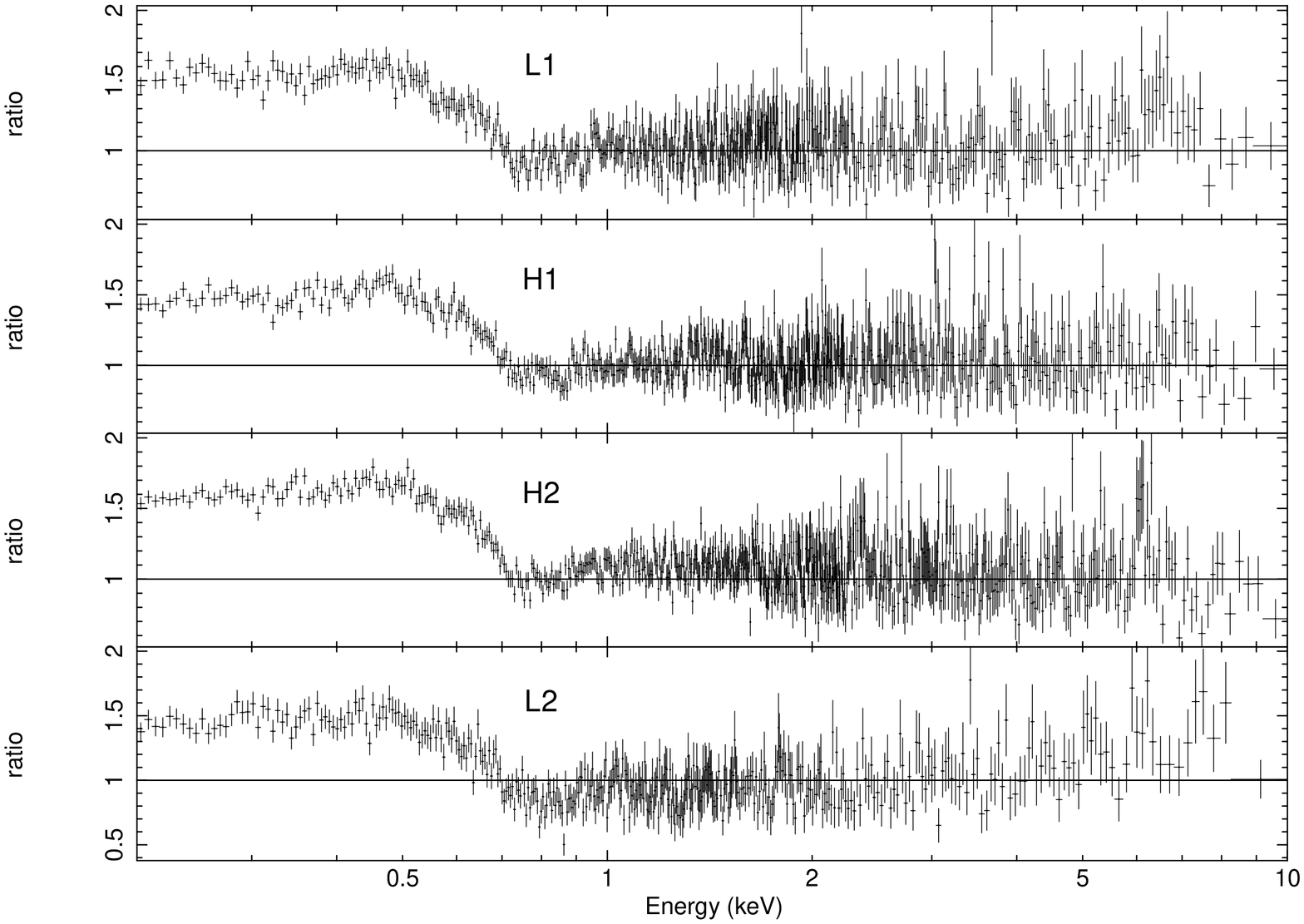}}
\end{minipage}
\caption{Left panel: the Obs1 spectrum of Was 61, the Galactic absorbed power-law model 
inferred from fitting in 2-10\keV band and its extrapolation to the soft X-ray band 
is also shown; Right panel: the ratio of data to model in each
state flagged as $L1$, $H1$, $H2$, and $L2$, comparing with the Obs1 spectrum.}
\label{fig:4statepo}
\end{figure}

We show the X-ray spectrum and best-fit power-law model in 2-10 keV, 
extrapolating to 0.2-10\keV band, for entire obs1 in the left panel of 
Figure~\ref{fig:4statepo}. The spectrum is approximately an average of 
$L1$, $H1$, and $H2$, considering roughly the same exposure time for 
each segment. By fixing the power-law index to 2.2--the value from the best 
fitted model of entire $obs1$ spectrum in 0.2-10\keV band--we re-normalized 
the model to fit the 2-10\keV spectra of $L1$, $H1$, $H2$, and 
$L2$. The ratios of data to the model for $L1$, $H1$, $H2$,
and $L2$ are shown in the right of Figure~\ref{fig:4statepo} to illustrate 
the variations of the X-ray spectrum. It is evident that power-law slope remains 
constant except for $L2$, for which 2-10\keV spectrum appears flat (see also Table
\ref{tab:poga1}). Soft excesses are prominent below 0.7\keV in all spectra.

Prominent variations of Fe K emission lines are seen on timescales 
as short as $\sim$20 ks. The broad Fe K$\alpha$ emission line is 
clearly present in $L1$ and may be present in $L2$, but it is 
not detected in other spectra. On the other hand in $H2$ there is a 
significant narrow Fe K$\alpha$, which is not detected in other spectra.
It is surprising that we observe such short-term Fe K$\alpha$ variations, 
especially the narrow component. We present a detailed 
analysis of variations of these spectral components in the subsequent
subsection.

\subsubsection{Variations of Fe K Emission and Continuum}

\begin{figure}[htbp]
\begin{minipage}{0.43\textwidth}
\centerline{\includegraphics[width=0.8\textwidth,angle=-90]{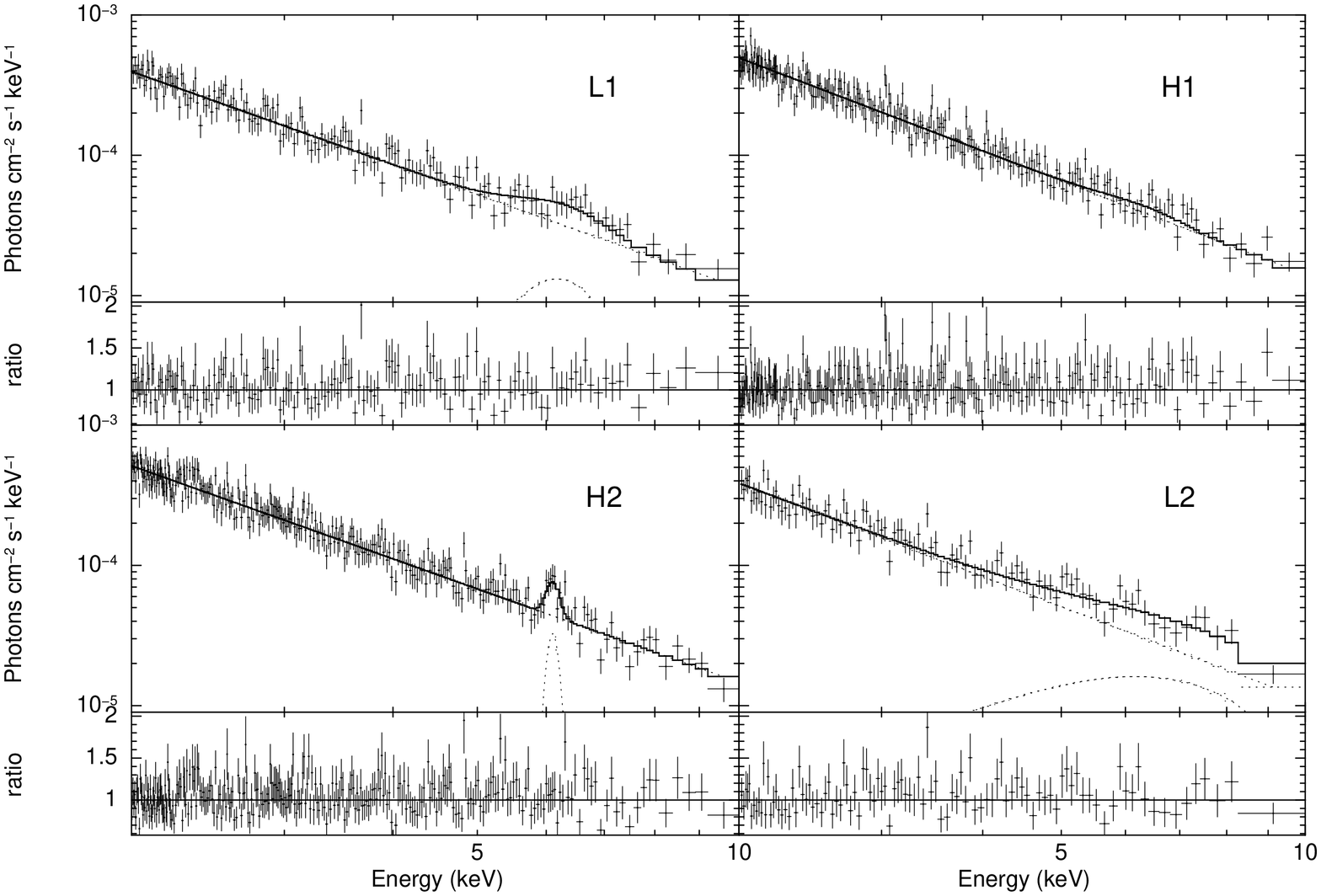}}
\end{minipage}
\hfill
\begin{minipage}{0.43\textwidth}
\centerline{\includegraphics[width=0.8\textwidth,angle=-90]{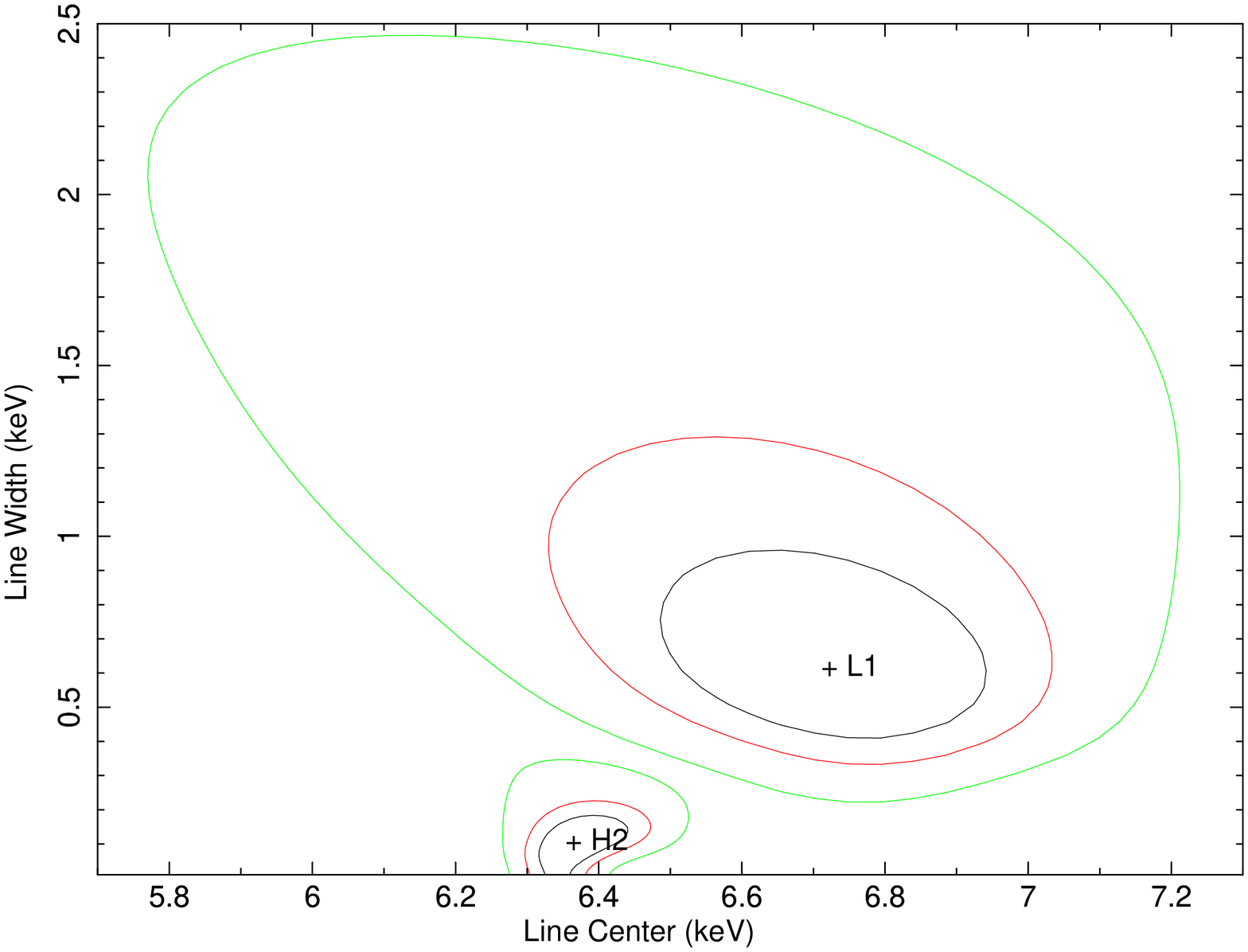}}
\end{minipage}
\caption{Left panel: the spectrum in 2-10~keV and the ratio of data to model 
in each state. The photon index of the Galactic absorbed power-law model is 
fixed at 2.2 and the fitted Gaussian line model are also shown. 
Right Panel: confidence contours (at the 68\%, 90\%, and 99\% confidence levels) 
of the line center and width of the Fe K line in $L1$ and $H2$.}
\label{fig:4poga}
\end{figure}

Because the spectrum in 2-10\keV is hardly affected by the warm 
absorption (see Figure~\ref{fig:spectr}), we initially fit each spectrum 
in the 2-10\keV using a simple power-law with the Galactic absorption. 
All fits are statistically acceptable (Table 1).  Because the systemic residuals 
around 6-7\keV are visible in $L1$ and $H2$, we add a Gaussian line 
for Fe K$\alpha$ to the above model. The fit is improved for $L1$ and $H2$ 
with $\Delta \chi^2=18.4$ and 23.4, respectively, using three more free 
parameters. According to $F-test$, the Fe K$\alpha$ line is significantly 
detected in $L1$ and $H2$ (left panel of Figure~\ref{fig:4poga}). However, 
the line widths and energies are significantly different (right panel 
of Figure~\ref{fig:4poga}). In $L1$, the line is broad ($\sigma=0.52\pm 
0.22$ keV) and likely comes from ionized gas ($E=6.74\pm$0.25 keV), while 
in $H2$, the line is much narrower ($\sigma$=0.11$\pm$0.07 keV) and 
the line center is at 6.37$\pm$0.06 keV. Adding a Gaussian line does
not lead to significant improvement for $L2$ ($P_{null}=78\%$) and 
$H1$ ($P_{null}=73\%$). 

 We check whether the $H1$ spectrum possesses the same broad 
line as $L1$, but the broad line is hidden in high continuum 
flux. We fit $L1$ and $H1$ spectra jointly with a power-law 
plus Gaussian model by tying the center, width, and normalization 
of the Gaussian component together. We get a total $\chi^{2}=358.5$ 
for 387 dof. We then allow the line normalizations to vary independently. 
This reduces $\chi^{2}$ by 6.1 for one more free parameter ($P_{null}
=1$\%). Thus, the broad line is weaker in $H1$ than in $L1$ at a 99$\%$ 
confidence level. To further assess the weakness of broad Fe K$\alpha$ 
in $H1$, we fix the line width and line center with the best-fit values 
for $L1$, and obtain the upper limit (at 90\% confidence level) of 
Gaussian normalization to be 53\% of that of $L1$. Because both $H1$ and $L1$ spectra 
are extracted from the first 40 ks exposure according to different 
count-rate, this result indicates that the broad Fe K$\alpha$ line responds 
negatively to the X-ray continuum flux.

We also check the presence of a similar broad Fe K$\alpha$ in 
$H2$ and $L2$ by fixing the line center and width to that of $L1$. We 
can only derive upper limits (at 90\% confidence level) on the Gaussian 
normalization for $H2$ and $L2$, which are 34\% and 90\% of the lower limit 
of $L1$. 

However, the $L2$ spectrum is consistent with the presence of a 
narrow line similar to that in $H2$. When the line center and width 
are fixed to the best-fit values in $H2$, the fit is significantly 
improved ($\Delta\chi^2$=3.8 for one more free parameter or $P_{null}
=3.8\%$) and the line flux is consistent with that of $H2$. The narrow 
Fe K$\alpha$ line is not detected in $L1$ and $H1$. We derive the upper 
limit (90\% confidence level) on the normalization of a narrow line 
with the same width and center as in $H2$ to be 45\% and 31\% of the 
lower limit of $H2$ for $L1$ and $H1$, respectively.

 Next, we check whether the variations in Fe K$\alpha$ are caused by 
imperfect modeling of the continuum. The power-law slopes are consistent 
to be same for $L1$, $H1$, and $H2$, so the variations of Fe K$\alpha$ 
are most likely not caused by continuum modeling among these spectra. 
However, $L2$ has a significantly flatter spectrum (Table~\ref{tab:poga1}) but 
its X-ray flux is even slightly higher than that in $L1$. It may be interpreted 
that the power-law slope only changes on relatively long timescales, as 
 suggested by Gardner~$\&$~Done~(2014) for the NLS1 galaxy PG~1244+026. 
Alternatively, the flatness of the $L2$ spectrum can be ascribed to 
a strong and very broad Fe K$\alpha$ line, or a strong reflection component. 
To check these possibilities, we apply some further exercises on the $L2$ spectrum. 
First, we fit an absorbed power-law model with a fixed photon index of 2.2, 
and obtain $\chi^2$=141.5 for 103 dof. Next we add a broad Gaussian to the model, 
and the fit is improved by $\Delta\chi^2$=50.3 for three more free parameters. 
The $\chi^2$ is almost the same as that of free-index power-law fit (Table 1), 
which yields a very broad ($\sigma$=2.1$^{+1.1}_{-0.6}$ keV) and unusual strong 
Fe K line (EW=2.7$^{+1.2}_{-0.8}$ keV). The equivalent width far exceeds 
theoretical predictions, thus the model is not favored.

However, the spectrum looks more likely to show some deficiencies at energies above 
8\keV, although with only one bin. This resembles an Fe K absorption 
edge in a reflection component. First, we apply a neutral reflection model 
({\it pexrav} in Xspec) to $L2$. The high energy cutoff is fixed at 
100 keV. The fit is acceptable with $\chi^{2}$=90.6 for 101 dof, and is 
comparable to the free power-law fit according to the F-test ($P_{null}$=42\%).
Next, we apply an ionized reflection model ({\it pexriv} in Xspec) for $L2$.
The best fit converges to an ionization parameter ($\xi$ $\geq$ 169 
erg~cm~s$^{-1}$) and is improved with respect to the free-index power-law 
model with $P_{null}$=5.6\% according to the $F-test$. As a result, we prefer to attribute 
the flat slope in $L2$ to the strong ionized reflection 
component, naturally with some Fe K$\alpha$ contribution.

We also check whether the broad line in $L1$ is due to a reflection component.
The fit with a neutral reflection model for $L1$ is significantly worse than 
that with the power-law plus broad Gaussian model ($P_{null}$=2.5\%), according 
to the F-test. Fitting an ionized reflection model to $L1$ spectrum results in 
a total of $\chi^{2}=148.9$ for 158 dof, which is comparable to the power-law plus 
broad Gaussian model ($P_{null}$=$87\%$). 
However, the fit gives a much steeper photon index of $\Gamma=2.5\pm0.3$ than 
those in other segments. Thus, the systematic residuals around 6-7 keV in $L1$ are 
most likely due to the broad Fe K$\alpha$ line. 

Because broad Fe K$\alpha$ is usually believed to be from an illuminated cold 
accretion disk, therefore, we use the relativistic accretion disk 
model to constrain the emission line region. 
We use the $diskline$ (Fabian et al. 1989) model in the $XSPEC$ for a disk 
surrounding a Schwarzshild BH. The parameters in the $diskline$ 
are the energy of the emission line, the inner and outer radii, inclination 
of the disk, the line emissivity index, and the equivalent width of the line. 
Considering the S/N of the spectrum around Fe K$\alpha$, 
it is not possible to constrain all of the model's parameters, so we assume 
an inclination of the disk to be 30$^o$, which is the average value for randomly inclined disks, 
and fix the emissivity index to $\beta=-3.0$ ($I(r)\propto r^{\beta}$), which is an 
expected value for a lamppost illuminating X-ray source at a height 
$h<<r$ above the symmetric axis of the disk. We also fix the outer 
disk radius to a large value of 800 $r_g$, beyond which there will be little 
Fe K$\alpha$ line emission from the disk for the given $\beta$; we 
leave the inner radius and the line flux as free 
parameters. With these assumptions, we obtain an inner radius for the 
disk of $\sim$6 $r_g$, with an upper limit of 31 $r_g$ for $L1$ and 
$<7$ $r_g$ for $L2$ at 90\% confidence level. Applying the same model to 
$H2$, we obtain an inner radius of $372_{-231}^{+428}$ $r_g$.
 We also use the $laor$ (Laor 1991) model in the $XSPEC$ for the disk 
surrounding a rotating BH. We assume $\beta$ and the inclination angle to be the same 
values as in $diskline$ model, but fix the outer disk radius to be 400 $r_g$ instead. 
The fit yields an inner radius consistent with that of a $diskline$ model 
within uncertainties. Thus, from $L1$ to $H2$, the inner radius increases 
by a factor of 10 or more if these lines are from a lamppost-illuminated 
accretion disk, and the line-equivalent width decreases by a 
factor of $\simeq$3.

\subsubsection{Variations of Soft Excess}
\label{sec:softvariation}

We use the blackbody model to characterize the soft excesses and fit the 
broadband spectra of $L1$, $H1$, $H2$, and $L2$ using the best model in \S \ref{full}. 
Because the 2-10 keV spectrum is consistent with a power-law index of 2.2 plus 
an additional Fe K$\alpha$ line or a reflection component in $L2$, we will 
fix the photon index of the power-law component to 2.2{\footnote{We 
also check if power-law slope is varying during the 0.2-10 keV fits. 
We jointly fit the spectra of $L1$, $H1$, and $H2$ using the best model in \S \ref{full}. 
Leaving the photon index varying free in $L1$, $H1$, and $H2$,
the fit result is consistent with that from the fixed the photon index to 
2.2 ($\Delta\chi^2=2.1$ for 3 more free parameters, $P_{null}=0.55$).}}. 
The parameters of 
the Fe K$\alpha$ line are fixed to the best-fit values derived from 
the fit to the hard X-ray spectrum with a power-law plus Gaussian line 
model for each segment because the hard X-ray spectrum is barely 
affected by the soft excess once the power-law index is fixed. As for 
the fit to the spectrum of entire obs1, we also include the warm 
absorption model as described in \S \ref{full}. 

\begin{figure}
\begin{minipage}{0.43\textwidth}
\centerline{\includegraphics[width=0.8\textwidth,angle=-90]{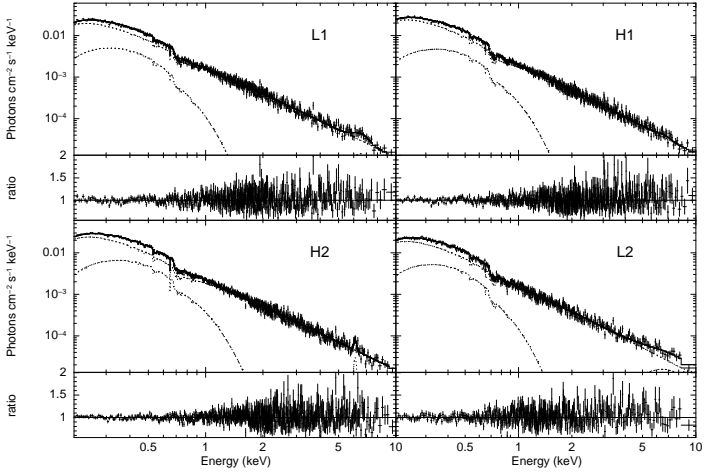}}
\end{minipage}
\hfill
\begin{minipage}{0.43\textwidth}
\centerline{\includegraphics[width=0.8\textwidth,angle=-90]{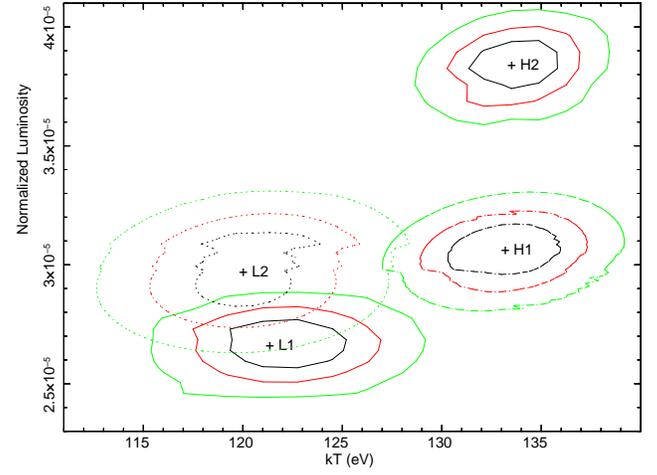}}
\end{minipage}
\caption{Left panel: spectrum and the ratio of data to model in each state;
 the model is a power-law plus a broad Gaussian component and a blackbody component,
which includes the Galactic absorption and a warm absorption.
Right panel: confidence contours (at the 68\%, 90\%, and 99\% confidence levels) of the
normalization and temperature of the blackbody model \label{fig:4state}}
\end{figure}

We find that warm absorption component does not vary significantly 
during the period of the {\it XMM-Newton} observations. Initially, we fit 
the spectra of $L1$, $H1$, $H2$, and $L2$ jointly by tying both the column density and 
ionization parameter of the absorber. Statistically the fit is acceptable 
(see the left panel of Figure~\ref{fig:4state} and Table~\ref{tab:bestmodel}). 
Then, we fit each spectrum independently. We find that the column density and 
ionization parameters of the warm absorber are consistent, and are also 
consisent with those derived from fitting the average spectrum of obs1.
In the following we fit the four spectra simultaneously, tying the 
parameters of the warm absorber, and examine the variability of the blackbody 
parameters. 

 We find that the blackbody component varies significantly. Initially, 
we tie all the normalizations and temperatures of blackbody components of 
$H1$, $H2$, and $L2$ to those of $L1$, and perform a joint fit. This results in 
a $\chi^{2}$/dof=2313.1/2154. When we separately untie the normalizations of 
$H1$, $H2$, and $L2$, $\chi^{2}$ is reduced by 4.3, 80.7, and 4.3, 
respectively. The fit is improved significantly according to 
the $F-test$ ($\Delta\chi^2=89.3$ for three more free parameters, 
$P_{null}=3\times10^{-18}$). Next, we separately untie the temperatures of $H1$, $H2$ and $L2$, 
which further reduces $\chi^{2}$ by 7.6, 28.0, and 0.3, respectively. The 
fit is improved at a significance of $P_{null}=1\times10^{-7}$. We show 
the confidence contours between the normalizations and temperatures of the 
blackbody components 
in each state (right panel of Figure~\ref{fig:4state}).
The temperatures of the soft excess are significantly higher (at $>$99\% 
confidence level) in $H1$ and $H2$ than in $L1$ or $L2$ (right panel of 
Figure~\ref{fig:4state}). The normalization of the blackbody is also 
significantly higher in $H2$ than in $H1$. The temperatures of the blackbody are 
consistent with being the same within their 68\% confidence contour for 
$L2$ and $L1$, while their normalizations differ at more than 68\% confidence level. 

We find some interesting trends between the variations of power-law 
and blackbody component on short and long timescales. We calculate 
the luminosity ratios of the blackbody and the power-law component 
in the 0.2-2\keV band, and find that they are similar in $L1$ (0.229$\pm$0.020) 
and $H1$ (0.225$\pm$0.018), but are significantly larger in $H2$ 
(0.262$\pm$0.039) and $L2$ (0.269$\pm$0.019). This suggests the possibility 
that on short timescales ($<$20 ks), the 
variations of soft excess and power-law components have the same 
amplitude, while on long timescales ($>20$ ks), the two are decoupled, 
by considering the fact that $L1$ and $H1$ are extracted in the intersected 
segments of the first 40 ks. Another interesting trend is that the 
blackbody temperature rises with increasing power-law flux (Table 
\ref{tab:bestmodel}), especially when the {\it ROSAT} result is considered. 

As commonly seen in other AGNs, the temperature of the blackbody is much higher 
than that expected in the innermost region of an optically thick and 
geometrically thin accretion disk. Heat advection in a slim disk (Abramowicz et al. 1988) can increase 
the disk temperature somewhat, but may not be able to fully resolve it. Furthermore, 
it predicts an accretion rate-dependent blackbody temperature, which is not 
consistent with observations (Done et al. 2012). Thus, the soft excesses cannot 
be the high energy tail of thermal disk emission. It was proposed that the soft 
X-ray excesses are formed via the reflection of X-rays by an ionized disk (e.g., 
Ross \& Fabian 2005) or through Comptonization process in the corona or the 
boundary layer between corona and accretion disk (e.g., Czerny et al. 
2003). In the following, we will examine these physical motivated models.

Firstly, we consider the disk reflection model. In this model, the accretion 
disk is partially ionized so the gas opacity is largest around 1 keV due to the
absorption edges of partially ionized O, Ne, Mg, Ar, Fe and so on. In addition, the
emission lines from the ionized gas cause a curvature in the reflected 
spectrum around 1 keV. Due to relativistic broadening, these edges and emission 
lines are blended to form a pseudo-continuum below 1 keV. One advantage 
of this model is that it accounts for the Fe K$\alpha$ line simultaneously. 
We use the ionized disk 
reflection model from Ross~\&~Fabian~(2005; {\it{reflionx}} in 
XSPEC), blurred by the $laor$ kernel ({\it kdblur} in XSPEC). The model has 
seven free parameters: photon index of ionizing continuum ($\Gamma$), the gas 
abundance ($Z$), the disk inner and outer radii ($r_{in}$ and $r_{out}$), disk 
inclination ($i$), ionization parameter of the disk atmosphere ($\xi_{r}$), and 
normalization of the reflection component. Together with primary power-law 
and previously identified warm absorber component, the final combined model 
has 11 free parameters. In the fitting, $\Gamma$ is tied to observed 
power-law continuum, assuming the same continuum is illuminating the disk. 
Further, we assume a solar abundance, and fix the outer radius of the 
disk to 400\, $r_g$, and the emissivity index of the disk to the standard 
value of 3. 

Considering the large number of free parameters, we first apply this model 
to the high S/N ratio spectrum of the entire obs1. The model does 
not yield an acceptable fit in the Fe K$\alpha$ region. We 
tried to free the metallicity, but this only marginally increased the 
Fe K$\alpha$ line. So we add a Gaussian line. The final model 
($\it{tbnew*warmabs*(powlaw+kdblur*reflionx+zgauss)}$) is marginally 
acceptable (Table 3). The best fit converges to a steeper $\Gamma$, relatively 
small inner disk radius, small ionization parameters, and a face-on viewing 
angle. The extra Fe K$\alpha$ line has higher energy and is narrower than in the 
fit with blackbody model. This is because the Fe K$\alpha$ associated with the 
reflection component has an intrinsic energy of 6.4 keV (low $\xi_r$) and is 
subject to gravitational redshift (face-on). However, despite the large number of 
free parameters, this fit is significantly worse than that with a blackbody 
model.   

Then we fit the model to the four segments of the spectra simultaneously. Following 
the results in the previous section, we tie $\Gamma$ and parameters of the warm 
absorber together. We also lock the disk inclination together. Because 
$R_{in}$ and $\xi_r$ are consistent for different segments, 
we also tie them together in the final fitting. The best fit for obs1 is significantly 
worse than the blackbody models (Table\,\ref{tab:reflmodel}).
Furthermore, considering the small inner radius, the ionization parameter appears to be
too low, while previous works found it much large $\xi$ (100-1000 erg cm s$^{-1}$, 
\eg, Ai et al. 2011, Laha et al. 2013) in other AGNs. 
So this model is less favored.

Next, we consider the Comptonization model. We use {\it optxagnf} in 
the $XSPEC$ (Done et al. 2012), taking into consideration thermal 
Comptonization in the inner accretion disk. The model consists of three 
components: (1) thermal disk emission from the outer disk ($r>r_{cor}$), 
(2) soft X-ray excess from the inner Comptonized warm disk 
($r_{iso}<r<r_{cor}$), and (3) power-law X-ray continuum from an optically 
thin, hot corona. One advantage is that the model covers the spectral 
energy distribution (SED) from optical UV to X-rays, simultaneously.
 However, we will not attempt to fit the SED from UV to X-ray, 
because of the uncertainties in the extinction correction to the OM flux. 
In addition, the OM exposures do not strictly match with those of the X-ray spectra, 
which is variable during the {\it XMM-Newton} observation. 
Instead we fit the X-ray spectra alone, then extrapolate the model to the 
UV to check its consistency with the average OM fluxes. As in previous 
fit, the final model ($\it{tbnew*warmabs*(optxagnf+zgauss)}$) also includes 
a Gaussian line and warm absorption, as well as the Galactic absorption.

\begin{figure}
\epsscale{.90}
\begin{minipage}{0.43\textwidth}
\centerline{\includegraphics[width=0.8\textwidth,angle=-90]{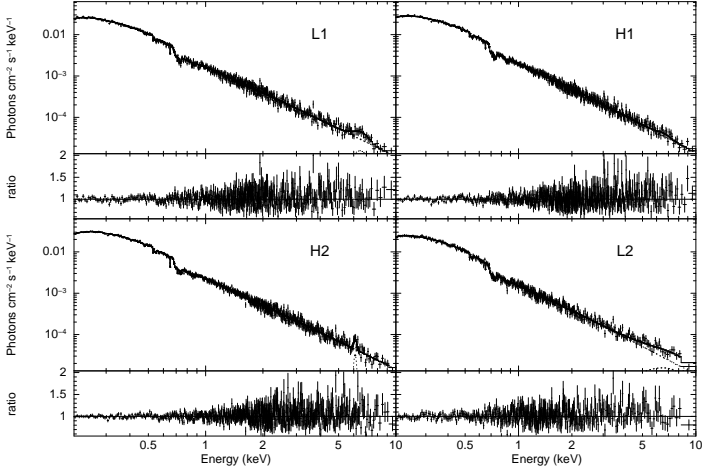}}
\end{minipage}
\hfill
\begin{minipage}{0.43\textwidth}
\centerline{\includegraphics[width=0.8\textwidth,angle=-90]{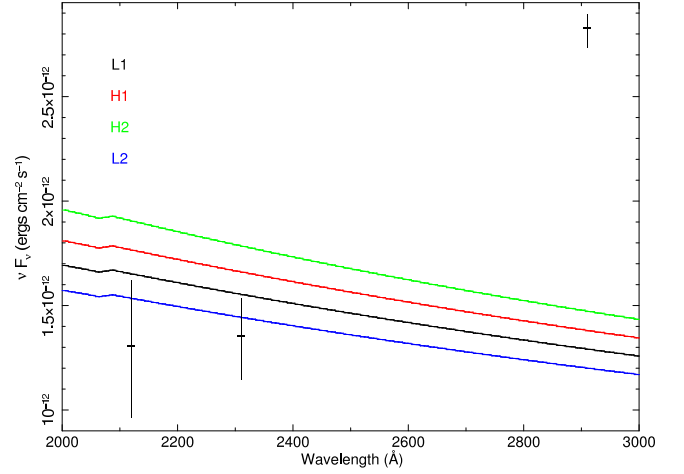}}
\end{minipage}
\caption{Left panel: spectrum and the ratio of data to the model 
{\it{tbnew*warmabs*(optagnf+zgauss)}} in each state; 
Right panel: extrapolation the model of each state to 2000-3000\AA, 
the symbols of the dots are the average fluxes of the OM
observations with filters $UVW1$, $UVM2$, and $UVW2$ (with 
statistical errors 3\%, 15\% and 25\%), respectively.
\label{fig:4optagn}}
\end{figure}

This model was first applied to the broadband spectrum of obs1. During 
the fitting process, we fix the BH mass to 4.6$\times$10$^{6}$ M$_{\sun}$, 
$\Gamma$ to 2.2, and Fe K$\alpha$ and warm absorption parameters 
to the best-fit values for model $\it{tbnew*warmabs*(powerlaw+zbbody+zgauss)}$ 
in Table \ref{tab:bestmodel}. Overall, the model gives as good a fit as 
the blackbody model with two more free parameters. Even with the high S/N spectrum, 
the coronal radius ($r_{cor}$) and electron scattering 
optical depth of the warm gas ($\tau$) remain poorly determined. Next we jointly fit this model 
to the spectra of four segments. The best-fit parameters of $optxagnf$ 
are listed in Table ~\ref{tab:comptmodel} and the fit is shown in Figure~\ref{fig:4optagn}. 
It is not surprising that the model also gives a good fit to the individual 
spectrum. The transition radius from a 
color corrected blackbody emission to an optically thick Comptonized disk are 
all around $20-60$ $r_g$, the electron scattering optical depth is $\tau=50-100$, 
and the electron temperature is around $kT=120-150$ eV. 

 We attempt to identify the dominated variable parameters among 
these spectra. After fixing the parameters of the Gaussian component, 
we tie $\tau$, the electron temperature for the soft Comptonization component 
$kT_{e}$, $r_{cor}$, and the fraction of the power inside $r_{cor}$, 
$f_{pl}$ first. We get $\chi^2/dof$=2229.3/2154 (see Table~\ref{tab:comptmodel}). 
We then untie the parameter $f_{pl}$ and the fit is 
significantly improved ($\Delta\chi^2=23.2$ for 3 dof, $P_{null}=5\times10^{-5}$). 
We mark this model as the base model in the further examinations. Next, 
we untie the parameter $kT_{e}$, or $\tau$, or $r_{cor}$, separately. In comparison with 
the base model, $\chi^2$ is reduced by 34.3, 35.1, or 22.0 for three dof, 
corresponding to $P_{null}=2\times10^{-7}$, 2$\times$10$^{-7}$, 
8$\times$10$^{-5}$, respectively. Apparently the fit depends equally
on $kT_e$ and $\tau$ and only slightly less on $f_{pl}$ and $r_{cor}$. 
Therefore, these parameters are highly degenerated.
 
This model also predicts the fluxes in the {\it UVW1} and {\it UVM2} bands in consistency 
with the observed {\it OM} flux if the intrinsic dust extinction is not important 
(Figure~\ref{fig:4optagn}). However, as we will discuss later, there is
good evidence that the Seyfert nucleus is reddened and these models 
give a far lower flux in {\it UVW2}. 

\section{Discussion}

The X-ray spectrum of Was 61 displayed most of the features observed in AGNs 
(i.e., a power-law component, warm absorption, soft X-ray excess, and Fe K$\alpha$ 
emission lines). We split the X-ray observation into four segments and 
characterized the variability of different spectral components in terms 
of empirical models. We found that all components but warm absorption were 
variable during the {\it XMM-Newton} observation. 

\subsection{Warm Absorber} \label{abs}

The absorber of Was~61 is well modeled by the $\it{warmabs}$ in the {\it pn} 
spectrum. 
The absorber has a relatively low ionization with $\xi \sim$ 6 erg cm s$^{-1}$, 
and a moderate column density N$_H\sim3.3\times 
10 ^{21}$ cm$^{-2}$. Short timescale variability of the warm absorber column 
density is not detected. A detailed modeling of the RGS spectrum suggests a slightly 
high ionization and column density and a super-solar abundance of N, but a 
sub-solar abundance of C and Fe.

The super-solar N and sub-solar C and Fe may indicate that C and 
Fe are depleted because C is an important composition of the various 
dust grains, such as graphite, carbonaceous particles, amorphous carbon particles, 
polycyclic aromatic hydrocarbon, and so on. Overabundance of N can be due to high gas metallicity 
because as a secondary element, the abundance of N varies with gas metallicity 
as $Z^2$; whereas primary elements, such as C and Fe, are proportional to 
$Z$. However, this cannot explain why C is significantly lower than O.

The absorber of Was~61 was also detected by {\it ROSAT} (Grupe et~al.~1999b). 
These authors fit the spectrum with a model consisting of a neutral absorption 
plus an absorption edge of O {\sc vii} at 0.74 keV, and derived a column density 
$N_{H}\sim 2.8\times10^{21}$~cm$^{-2}$. We refit the {\it ROSAT} spectrum with 
the same warm absorption model for the {\it XMM-Newton} spectrum, and obtain an 
$N_{H}$, which is consistent with that from {\it XMM-Newton} data and Grupe et al., but 
with a lower $\xi < 2$ erg cm s$^{-1}$ than that from {\it XMM-Newton} data. 
The result can be explained as the same absorbing material exposed to an 
ionizing continuum that is weaker than that of the {\it XMM-Newton} data. 
This is consistent with the result that the power-law continuum in the 
{\it ROSAT} spectrum is a factor of 2.6 lower.

Since the change in the ionization parameters occurred between 1991 December 
and 2005 June, we set the upper limit of the recombination timescale 
to be 13.0 year (in the rest frame of the source). Using the recombination 
timescale given by Blustin et al. (2005) at the ionization equilibrium, 
we estimate the lower limit of the electron density to be n$_{e}$ $\sim$ 10$^{3}$ cm$^{-3}$, 
assuming the absorber is dominated by O {\sc vii} and/or O {\sc viii}, and the 
electron temperature is $T_{e}\sim10^{5}$ K, which is typical for a photon-ionized gas. 
Combined with observed $N_H$ and $\xi$, we derive the upper limit of the thickness of 
the absorber to be $\sim 1 \pc$, and an upper limit of the distance from 
the absorber to the AGN to be $\sim50\pc$.

Was~61 is an infrared luminous source (Keel et al. 1988) and its optical spectrum 
shows a large reddening in both the continuum and emission lines. Du et~al.~(2014) 
obtained Balmer decrements H$\alpha$/H$\beta$=5.71 and 6.14 for the broad and 
narrow lines, respectively. These Balmer 
decrements correspond to a color excess $E(B-V)\sim$ 0.6 (Xiao~et~al.~2012) 
assuming an intrinsic H$\alpha$/H$\beta$=3.1 and average Galactic extinction 
curve. The similar Balmer decrements for broad and narrow lines indicate that 
the absorber is located outside the NLR. 
 
The depletion of both C and Fe indicates that the dust is within the warm 
absorber, although the detection of the grain signature (such as Fe L2 and L3 edges) 
will be crucial to verify such a scenario (Lee et al. 2001; 2009; c.f. Sako et al. 
2003). Note that the upper limit on the distance of the warm absorber is consistent 
with the above dust location. 
If the same material is responsible for both the X-ray absorption and optical reddening, 
we can estimate the dust-to-gas ratio $E(B-V)/N_H$ $\sim$ 1.9$\times$10$^{-22}$ mag 
cm$^{2}$. This value of dust-to-gas ratio is consistent with that for the Milky Way 
(Predehl \& Schmitt 1995; Nowak et al. 2012). This fortuitous agreement also indicates 
that dust is within the absorber.  
 
\subsection{On the Fe $K\alpha$ Line} \label{soft}

We find that the Fe K$\alpha$ line is highly variable. When the line is 
described empirically in Gaussian, the width, centroid energy, and 
flux vary significantly on timescales of about 20 ks. In the low flux 
segment ($L1$) of the first 40 ks of obs1, the line is broad and strong, 
while in the high flux segment ($H1$) of the same period the broad Fe K$\alpha$ 
flux is a factor of at least 2 times lower. In the later segments, $H2$ 
and $L2$, we only detect a narrow Fe K$\alpha$ line. This narrow Fe K$\alpha$ 
is not detected in the first 40 ks observation.

Variations in the Fe K$\alpha$ line can be caused by changing accretion 
mode, the ionization of disk atmosphere, or the geometry and kinematic of 
the X-ray emission region. Firstly, we cannot associate the variation 
between $L1$ and $H1$ to the transition in the accretion mode because the 
integration time of $H1$ and $L1$ is alternative, and also the time interval 
is shorter than the viscosity at the typical line formation radius, 10 $r_g$. 
Secondly, the variation of the X-ray continuum between $L1$ and $H1$ is only moderate 
(20\%). It should not give rise to a sufficiently large change in disk 
ionization, which would dramatically affect the X-ray emission. Thus, 
we ascribe the decrease in the line flux in $H1$ to the variation of anisotropy 
X-ray emission. If the X-ray emission is highly beamed toward the vertical 
direction during $H1$ while it is isotropic during $L1$, then the disk 
will see a lower X-ray continuum than us, resulting in a weak iron 
K$\alpha$ line and reflection component. Anisotropic X-ray continuum 
emission was suggested based on the correlation between the fluxes of hard 
X-ray and infrared high ionization lines (Liu et al. 2014), and a change 
in the corona structure or kinematics was inferred from a recent 
analysis of the variability of reflection components (Wilkins \& 
Fabian 2012; Gallo et al. 2015).

The presence of a narrow Fe K$\alpha$ line and lack of a broad Fe K$\alpha$ line 
in $H2$ may be explained in the same framework. A relativistically outflowing 
corona or extended corona will weaken the illumination in the inner disk 
much more than in the outer disks, resulting in a flat line emissivity 
toward large radii (Wilkins \& Fabian 2012), thus a narrower line. 
In the diskline fit to $H2$, the inner disk radius is constrained 
to larger than 100 $r_g$. This requires either a very extended corona or
a combination with relativistically beaming. 

Drastic variations in line width on such a short timescale have been 
reported only occasionally (\eg, Petrucci et~al.~2002). In Mrk 841, the 
narrowness of the line is interpreted as arising from a locally 
illuminating spot in the inner accretion disk by a flare near the 
disk surface in a snapshot (Petrucci et~al.~2002). In this scenario, 
the line center shifts due to the Doppler effect of the illuminated gas as the 
hot spot rotates (Longinotti et~al.~2004). The orbital timescale 
t$_{k}=2.85 \times M_{6} (r/20r_{g})^{3/2}$ ks (r$_{g}=GM_{BH}/c^{2}$) 
is relevant for such shift, assuming a corotating hot spot (Treves et 
al. 1988). This means that the observed line energy 6.4 keV is most likely a Fe K$\alpha$ line.
If the integration time is a significant fraction of 
the orbit timescale, the line will be broadened due to the superposition 
of the line at different times. The narrowness of the line will require 
the integration time (20 ks for $H2$) to be much shorter than the orbit 
period. For a BH mass of Was 61, this translates to a line-emitting 
region of larger than 100 $r_g$. On the other hand, if the disk is seen 
nearly face-on, the Doppler broadening would be insignificant, which means that the 
hot spot can be close to the BH. However, the line is redshifted due 
to gravitational redshift; the observed line energy 6.4 keV limits 
the radius of the line emission line region to $>14 r_g$ for a 
Schwarzschild BH and a H-like iron. 
 
 \subsection{On the Soft X-Ray Excess}
 
We characterized the soft excesses in terms of an empirical 
blackbody model and found that both normalizations and
temperatures were varying during the {\it XMM-Newton} observation, 
and between {\it ROSAT} and {\it XMM-Newton}. The temperature of 
the blackbody appears to increase with the flux of a power-law component. 
This suggests that the origin of the soft excess is closely related
to a power-law component. We also find a tentative trend that the 
flux ratio of the blackbody to power-law vary on long timescale 
($>20 ks$), but not on short timescales. This behavior is expected 
in thermal Comptonization model for hard X-rays if the soft X-rays are 
the seeding photons. The power of Comptonized light is proportional to 
the strength of the input soft X-rays on timescales shorter than that of the structure 
change of corona, while on long timescales, 
electron temperature or/and optical depth may vary and the proportionality 
is destroyed. A complete analysis should decomposite the X-ray photons 
into different Fourier components, which requires a longer exposure time.

The blackbody does not provide a physical model for soft X-ray excess 
because the derived temperature is much higher than the expected value 
in the inner accretion disk. We also applied two physical models: ionized 
reflection and a Comptonization model for the soft excess in subsection~
\ref{sec:softvariation}. We also find that an ionized reflection model does not 
give a good fit to the broadband X-ray spectrum for the entire obs1 
or for individual segments. In particular, the model does not reproduce 
the strength of Fe K$\alpha$, which is very different from results for 
other AGNs (e.g. 1H 0707-495, Fabian et al. 2009, Zoghbi et al. 2010; 
RE J1034+396, Zoghbi \& Fabian 2011). 
Even with an additional Fe K$\alpha$ added, the final 
fit is still much worse than that with the blackbody model. In addition, 
the ionization parameter of the disk is a factor of 10-100 lower than 
those found in other AGNs, while the model requires the disk to extend 
down to a few gravitational radii. Thus the reflection model is not favored 
here. 

The Comptonization model {\it optxagnf} can reproduce the observed spectrum 
of either entire obs1 or an individual segment. The model gives an $L/L_{Edd}
=0.22-0.30$, which is within the upper limit of a geometrically thin disk model. The 
corona radius is in the range of $20-60 r_g$, and the temperature ($kT$) and optical 
depth of the warm media are $120-150$ eV and 50-100, respectively, for the 
spectrum of the individual segment. These parameters are highly degenerated 
and we are unable to isolate the major factor for the variations of soft 
X-ray excess.

\section{Summary and Conclusions}

The {\it XMM-Newton} X-ray spectrum of Was~61 can be well modeled with a power-law 
absorbed by low ionization gas plus a blackbody soft excess and a broad or 
narrow Gaussian Fe K$\alpha$ line. The X-ray flux varies significantly 
on timescale of ks during the {\it XMM-Newton} observation. 
Combined with the variability from the X-ray lightcurve with the temporal spectral analysis, 
we find that:

\begin{itemize}
\item The photon index of the power-law component remains constant 
($\Gamma\sim 2.2$) during the {\it XMM-Newton} observation, whereas the soft X-ray 
excess varies both in shape and normalization. We find tentative 
evidence that the blackbody temperature increases as the power-law continuum 
brightens, and the ratio of the blackbody to power-law component remains 
constant on short timescale but varies on long timescales. 

\item The absorber has a super-solar abundance of N, but a sub-solar 
abundance of C and Fe, a low ionization parameter, and it remains very stable 
during the {\it XMM-Newton} observation and possibly between the {\it XMM-Newton} 
and {\it ROSAT} observation 13 years ago. The change in the ionization parameter 
between {\it XMM-Newton} and {\it ROSAT} can be fully explained as being due to a photoionization
effect. From this variation, we set an upper limit of 50~pc on the distance to the 
AGN. The depletion of C and Fe relative to N in the warm absorber {\bf indicates} that 
the same absorber causes the reddening of both the broad and narrow emission lines. 
We derive a dust-to-gas ratio similar to the Galactic one. 
  
\item Fe K$\alpha$ varies dramatically during the {\it XMM-Newton} observation. 
A strong broad Fe K emission is detected in the low state ($L1$) of 40 ks exposure, 
but not in the high state ($H1$) of the same interval where the upper limit 
of the line flux is a factor of two weaker. We detect only narrow Fe~K emission 
in the remaining 
observation ($H2$ and $L2$). The same narrow line is not required by the data 
in the first 40 ks exposure. We interpret these variations as being induced by 
the change in the geometry and kinematic of the X-ray emission region, although 
the detailed physical process is still not clear.

\item The soft excess with a temperature ${\it{kT}}\sim 0.11-0.14$\keV 
can be better described with 
the model of the Comptonization of disc photons via a warm and optical thick 
inner disk, rather than the reflection from a partially ionized accretion disk. 
However, we are unable to identify the main factor driving the variability of 
the soft X-ray excess.

\end{itemize}

\acknowledgments
{We would like to thank the anonymous referee for the
suggestions and comments that greatly improved the paper.
This work is based on observations obtained with {\it XMM-Newton}, an ESA science 
mission with instruments and contributions directly funded by ESA Member States 
and the United States (NASA). We thank the {\it XMM-Newton} Operations, Software and Calibration 
teams. This work is supported by the Strategic Priority Research Program ``The 
Emergence of Cosmological Structures'' of the Chinese Academy of Sciences 
(XDB09000000), National Basic Research Program of China (grant No. 2015CB857005), 
and NSFC through (NSFC-11103071, NSFC-11233002, and NSFC-11421303).}

\clearpage

\begin{table}
\centering\scriptsize 
\caption{ Results of Fits to 2-10\keV Spectra \label{tab:poga1}}
\begin{threeparttable}
\begin{tabular}{c cc  ccc  c}\\
\hline
\hline
\multicolumn{4}{c}{Model: tbnew*(Powerlaw)} \\
\hline
&  \multicolumn{2}{c}{Powerlaw}   &        \\
\hline
&$\Gamma$  &  $L^{43}_{2-10}$ &  $\chi^{2}/dof$   \\
& (1) & (2) & (6)    \\
\hline
$obs1_{pn}$ & $2.13\pm0.03$  &  $1.95\pm0.03$  &   497.9/520 \\
\hline
$L1$ & $2.04\pm0.07$ &  $1.67\pm0.13$ & 165.1/160 \\
$H1$ & $2.12\pm0.05$ &  $2.00\pm0.13$ & 207.4/232 \\
$H2$ & $2.17\pm0.05$ &  $2.05\pm0.03$ & 302.1/278 \\
$L2$ & $1.84\pm0.07$ &  $1.84\pm0.18$ & 90.8/102  \\
\hline\hline
\multicolumn{7}{c}{Model: tbnew*(Powerlaw + zgauss)} \\
\hline
&  \multicolumn{2}{c}{Powerlaw}  &   \multicolumn{3}{c}{Gauss} &        \\
\hline
&$\Gamma$  &  $L^{43}_{2-10}$ &  $E_{center}$ & $\sigma$  &  EW &  $\chi^{2}/dof$   \\
& (1) & (2) & (3) &(4) &(5) & (6)   \\
\hline
$obs1_{pn}$ & $2.17\pm0.03$  &  $1.91\pm0.03$ &  $6.43_{-0.28}^{+0.25}$ & $0.49_{-0.26}^{+0.32}$ & $0.22_{-0.03}^{+0.02}$ &  480.0/517 \\
\hline
$L1$ & $2.14\pm0.08$ &  $1.60\pm0.15$ & $6.71_{-0.22}^{+0.28}$ & $0.52_{-0.26}^{+0.32}$  & $0.54^{+0.35}_{-0.26}$ & 146.7/157 \\
$H1$ & $2.15\pm0.06$ &  $1.98\pm0.14$ & 6.71(f) &              0.52(f) &                          $0.11^{+0.14}_{-0.11}$ & 206.0/231 \\
$H2$ & $2.21\pm0.05$ &  $2.02\pm0.03$ & $6.37_{-0.05}^{+0.06}$ & $0.11_{-0.07}^{+0.06}$   & $0.19^{+0.06}_{-0.06}$ & 278.7/276 \\
$L2$ & $1.88\pm0.10$ &  $1.81\pm0.22$ & 6.71(f) & 0.52(f) &     $0.13^{+0.22}_{-0.13}$ & 89.7/101  \\
$L2$ & $1.87\pm0.08$ &  $1.81\pm0.22$ & 6.37(f) & 0.11(f) &     $0.11^{+0.09}_{-0.09}$ & 87.0/101  \\
\hline
$L1$ & 2.2(f) &  $1.56\pm0.04$ & $6.73_{-0.27}^{+0.24}$ & $0.62_{-0.22}^{+0.40}$  & $0.74^{+0.28}_{-0.24}$ & 148.0/158 \\
$H1$ & 2.2(f) &  $1.94\pm0.04$ & 6.73(f) &              0.62(f)&       $0.22^{+0.14}_{-0.15}$ & 208.4/232 \\
$H2$ & 2.2(f) &  $2.02\pm0.03$ & $6.37_{-0.06}^{+0.08}$ &  $0.11_{-0.08}^{+0.07}$   & $0.19^{+0.06}_{-0.06}$ & 278.7/277 \\
$L2$ & 2.2(f) &  $1.52\pm0.07$ & 6.73(f) &              $2.08_{-0.56}^{+1.12}$ &   $2.71^{+1.18}_{-0.80}$ & 91.2/102  \\
\hline\hline
&\multicolumn{3}{c}{Model: $\it{tbnew*(Powlaw+Diskline)}$, $\Gamma=2.2$ } \\
\hline
& \multicolumn{2}{c} {$\it{Diskline}$} & \\ 
\hline
&  $E_{center}$  & $R_{in}$ & $\chi^2/dof$ \\ 
& (7) & (8) & (6)  \\ 
\hline
L1 & $6.81^{+0.58}_{-0.18}$ & $6.2^{+24.8}_{-*}$  &  152.5/158 \\ 
H2 & $6.40^{+0.10}_{-0.07}$   &  $371.7_{-233.2}^{+428.2}$ & 279.2/276 \\ 
\hline\hline
\multicolumn{6}{c}{Model: tbnew*pexrav} \\
\hline
&  \multicolumn{4}{c}{pexrav} &       \\
\hline
&$\Gamma$  &    $E_{cutoff}$ & $rel_{refl}$  &  $\chi^{2}/dof$   \\
& (9) & (10) & (11) & (6)    \\
\hline
$L1$ & $2.49_{-0.26}^{+0.35}$  & 100(f) &              $5.5_{-3.3}^{+6.9}$ &  153.7/159  \\
$L1$ & 2.2(f)  & 100(f) &              $2.0_{-0.6}^{+0.6}$ &  157.2/160  \\
$L2$ & $1.88_{-0.13}^{+0.37}$  & 100(f) &              $0.6_{-*}^{+3.9}$ &  90.6/101  \\
$L2$ & 2.2(f)  & 100(f) &              $3.9_{-0.9}^{+1.0}$ &  92.7/102  \\
\hline\hline
\multicolumn{7}{c}{Model: tbnew*pexriv} \\
\hline
&  \multicolumn{5}{c}{pexriv} &       \\
\hline
&$\Gamma$  &    $E_{cutoff}$ & $rel_{refl}$  &  $\xi$  & $\chi^{2}/dof$   \\
& (9) & (10) & (11) & (12) & (6)    \\
\hline
$L1$ & $2.49_{-0.23}^{+0.31}$  & 100(f) &              $4.2_{-2.3}^{+5.1}$ & $218.6_{-218.1}^{+1528.9}$ & 148.9/158  \\
$L1$ & 2.2(f)  & 100(f) &              $1.5_{-0.5}^{+0.7}$ &  $149.5_{-149.4}^{+1104.0}$ &   153.3/159  \\
$L2$ & $1.98_{-0.17}^{+0.22}$  & 100(f) &              $0.9_{-0.5}^{+1.5}$ & $4999.2_{-4830.1}^{+*}$ & 86.13/100  \\
$L2$ & 2.2(f)  & 100(f) &              $2.3_{-0.6}^{+0.5}$ & $4999.8_{-4889.9}^{+*}$ &  88.7/101  \\
\hline\hline
\end{tabular}
\begin{tablenotes}
\item[] The columns are (1)-(2), the photon index $\Gamma$, the 2-10 keV luminosity in rest frame (in unit 10$^{43}$ erg s$^{-1}$) with
the absorbing column removed of the power-law component;
(3)-(5) the energy line center (in unit keV), the one $\sigma$ line width (in unit keV), and the equivalent width (in unit keV) of the Gaussian line component;
(6) the $\chi^2$ and degree of freedom;
(7)-(8) the energy line center in the rest frame (in unit keV) and the inner radius (in unit GM/c$^2$) of the {\it diskline} model;
(9)-(11) the first power-law photon index, the cutoff energy (keV) and the reflection scaling factor of the {\it pexrav} or {\it pexriv} model;
(12) the disk ionization parameter of the {\it pexriv} model (in unit erg cm s$^{-1}$).
The uncertainties are given at 90\% confidence levels for one interesting parameter.
\end{tablenotes}
\end{threeparttable}
\end{table}

\begin{sidewaystable}
\caption{ Results of Broadband Spectral Fits for blackbody Models \label{tab:bestmodel}}\scriptsize 
\begin{threeparttable}
\begin{tabular}{c ccc ccc cc cc c}\\
\hline\hline
\multicolumn{12}{c}{Model: $\it{tbnew*warmabs*(powerlaw+zbbody+zgauss)}$} \\
\hline
 & \multicolumn{3}{c}{Powerlaw}  &  \multicolumn{2}{c}{Blackbody} &  \multicolumn{3}{c}{zgauss} &  \multicolumn{2}{c}{warmabs}  &     \\
\hline
&$\Gamma$ &  $L^{43}_{0.2-2}$  &  $L^{43}_{2-10}$ &  $\it{kT}$  & $L^{43}_{0.2-2}$ & E$_{center}$ & $\sigma$ &  EW & $N_{H}^{21}$ & $\xi$   &  $\chi^{2}/dof$   \\
& (1) & (2) & (3) &(4) &(5) & (6) & (7) & (8) & (9) &(10) &(11)     \\
\hline
$obs1_{pn}$ & $2.21\pm0.02$ & $4.13\pm0.05$ &  $1.91\pm0.02$ & $135.1_{-4.3}^{+2.0}$ & $1.19_{-0.20}^{+0.07}$ &  $6.41_{-0.22}^{+0.27}$ & $0.63_{-0.21}^{+0.22}$ & $0.25_{-0.04}^{+0.07}$ &  $3.31_{-0.24}^{+0.12}$  & $6.8_{-1.9}^{+0.66}$  &  927.4/875 \\
$obs1_{RGS}$ & 2.2(f) &\nodata &\nodata & 135(f) & \nodata&  \nodata& \nodata& \nodata& $4.48_{-0.22}^{+0.40}$ & $9.40_{-1.99}^{+1.11}  $  & 1142.7/945 \\
${\it ROSAT}$ & 2.2(f) & $1.57\pm0.04$  &\nodata & $102.2_{-4.6}^{+3.1}$ & $1.24\pm0.04$ &  \nodata& \nodata& \nodata& $2.72_{-0.68}^{+0.85} $ & $ < 2$   & 152.0/123 \\
\hline
$L1$ & 2.2(f) & $3.49\pm0.07$ & $1.61\pm0.03$ & $121.6_{-3.4}^{+4.2}$ & $0.80\pm0.04$  &  $6.71_{-0.22}^{+0.20}$ & $0.58_{-0.13}^{+0.27}$ & $0.66\pm0.20$ &   &  &     \\
$H1$ & 2.2(f) & $4.22\pm0.05$ & $1.95\pm0.03$ & $133.3_{-2.9}^{+2.0}$ & $0.95\pm0.03$  & 6.71(t) & 0.58(t) &                               $0.19\pm0.14$ &   &  &       \\
$H2$ & 2.2(f) & $4.53\pm0.05$ & $2.09\pm0.03$ & $134.2_{-3.1}^{+2.9}$ & $1.19\pm0.05$  & $6.37_{-0.06}^{+0.06}$ & $0.11_{-0.07}^{+0.06}$ & $0.17\pm0.06$ &  $3.20\pm0.16$ & $6.06^{+1.97}_{-0.80}$  & 2183.9/2142 \\
$L2$ & 2.2(f) & $3.30\pm0.07$ & $1.52\pm0.03$ & $120.1_{-3.4}^{+4.2}$ & $0.89\pm0.04$  & 6.71(t) & $2.27_{-0.52}^{+0.57}$ &       $2.86_{-0.71}^{+0.99}$ &  &  &     \\
\hline
\end{tabular}
\begin{tablenotes}
\item[] The columns are (1)-(3) the photon index $\Gamma$, the 0.2-2 and 2-10 keV luminosity in rest frame (in unit 10$^{43}$ erg s$^{-1}$) with
the absorbing column removed of the power-law component; (4)-(5) the temperture (in unit eV) and the 0.2-2 keV
luminosity in rest frame (in unit 10$^{43}$ erg s$^{-1}$) with the absorbing column removed of the blackbody component;
(6)-(8) the energy line center (in unit keV), the one $\sigma$ line width (in unit keV), and the equivalent width (in unit keV) of the Gaussian line component;
(9)-(10) the Hydrogen column density (in unit 10$^{21}$ cm$^{-2}$) and the ionization parameter $\xi$ ($\xi$=L/nR$^2$, in unit erg cm s$^{-1}$, see Done et al. 1992) of the warm absorber; (11) the $\chi^2$ and degree of freedom.
\end{tablenotes}
\end{threeparttable}
\end{sidewaystable}

\begin{sidewaystable}
\caption{{Results of Broadband Spectral Fits for reflection Models} \label{tab:reflmodel}}\scriptsize 
\begin{threeparttable}
\begin{tabular}{c cc cccc ccc cc c}\\
\hline\hline
\multicolumn{12}{c}{Model: $tbnew*warmabs*(powerlaw + kdblur*reflionx + zgauss)$}  \\
\hline
&  \multicolumn{2}{c}{Powerlaw}  & \multicolumn{4}{c}{kdblur*reflionx}  & \multicolumn{3}{c}{zgauss} & \multicolumn{2}{c}{warmabs} &    \\
\hline
&$\Gamma$ &  norm1 & $R_{in}$ & Incl & $\xi_{r}$ & norm2    & E$_{center}$ & $\sigma$ & norm3  & $N_{H}^{21}$ & $\xi$     &  $\chi^{2}$ \\
&         & 10$^{-3}$ &       &      &           & $10^{-6}$&              &          & 10$^{-6}$& $10^{21}$  &           &             \\
& (1) & (2) & (3) &(4) &(5) & (6) & (7) & (8) & (9) &(10) & (11) & (12)    \\
\hline
$obs1_{pn}$ & $2.33\pm0.01$ & $2.46\pm0.02$ & $4.41^{+0.28}_{-0.56}$ & $0.1^{+0.1}_{-0.1}$ & $11.9^{+0.9}_{-0.7}$ & $4.5^{+0.3}_{-0.2}$ & $6.68_{-0.19}^{+0.20}$ & $0.34_{-0.14}^{+0.23}$ & $6.6^{+3.4}_{-2.7}$ &  $2.8\pm0.1$ & $11.3^{+2.3}_{-0.6}$ & 971.4/873 \\
\hline
$L1$ &  & $2.02\pm0.02$ &  &  &  & $4.6^{+0.3}_{-0.2}$ & $6.84_{-0.23}^{+0.18}$ & $0.56_{-0.21}^{+0.19}$ & $19.3^{+0.7}_{-0.5}$ &      &  &  \\
$H1$ & $2.33\pm0.01$ & $2.48\pm0.02$ & $3.28^{+0.47}_{-0.31}$ & $21.1^{+3.3}_{-5.0}$ & $10.4^{+0.4}_{-*}$ & $5.1^{+0.2}_{-0.3}$ & 6.84(t) & 0.56(t) & $7.2^{+6.9}_{-4.2}$ &  $2.8\pm0.1$ & $12.9^{+0.2}_{-2.2}$ & 2283.9/2139\\
$H2$ &  & $2.69\pm0.02$ &  &  &  & $6.5^{+0.2}_{-0.3}$ & $6.36_{-0.06}^{+0.04}$ & $0.04_{-0.04}^{+0.11}$ & $5.5^{+2.3}_{-2.5}$   &  &  &   \\
$L2$ &  & $1.91\pm0.02$ &  &  &  & $5.6^{+0.4}_{-0.3}$ & 6.84(t) & $2.28_{-0.51}^{+0.81}$ & $78.0^{+18.7}_{-15.9}$ &   &  &   \\
\hline
\end{tabular}
\begin{tablenotes}
\item[] The columns are: (1)-(2) the photon index $\Gamma$, the normazation in 10$^{-3}$ photons keV$^{-1}$ cm$^{-2}$ s$^{-1}$ at 1 keV of the power-law component;
(3)-(6) the inner radius (in unit GM/c$^{2}$), the inclination (in unit degree), the ionization parameter $\xi_{r}$ ($\xi_{r}$=F/nR$^2$, in unit erg cm s$^{-1}$),
the normalization ($\times$10$^{-6}$) of the relativistic blurred reflected spectrum;
(7)-(9) the energy line center (in unit keV), the one $\sigma$ line width (in unit keV), and total 10$^{-6}$ photons cm$^{-2}$ s$^{-1}$ of the Gaussian line component;
(10)-(11) the Hydrogen column density (in unit $10^{21}$ cm$^{-2}$), and the ionization parameter $\xi$; (12) the $\chi^2$ and degree of freedom.
\end{tablenotes}
\end{threeparttable}
\end{sidewaystable}

\begin{table}
\centering
\caption{ Results of Broadband Spectral Fits for Comptonization Models \label{tab:comptmodel}}\scriptsize 
\begin{threeparttable}
\begin{tabular}{c ccccc c}\\
\hline\hline
\multicolumn{7}{c}{Model: $\it{tbnew*warmabs*(optxagnf+zgauss)}$, $\Gamma=2.2$, $M_{BH}=4.6\times10^{6}$ M$_{\sun}$ }   \\
\hline
&  \multicolumn{5}{c}{optxagnf}   &     \\
\hline
& $r_{cor}$ & $\tau$ & $\it{kT}$ & $f_{pl}$ & $ L/L_{Edd}$ &   $\chi^{2}/dof$    \\
& (1) & (2) & (3) &(4) &(5) & (6)    \\
\hline
$obs1_{pn}$ & $24.7_{-0.5}^{+0.5}$  & $88.7^{+*}_{-31.6}$ & $135.1_{-2.9}^{+6.1}$   & $0.90_{-0.01}^{+0.01}$ & $0.27\pm0.01$ & 929.8/879\\
\hline
$L1$ & $21.4_{-0.8}^{+0.8}$  & $51.8^{+10.3}_{-6.4}$ & $136.2_{-5.1}^{+4.8}$   & $0.89_{-0.02}^{+0.02}$ & $0.24\pm0.02$ &   \\
$H1$ & $27.3_{-0.6}^{+1.1}$ & $100^{+*}_{-23.7}$  & $135.5_{-3.4}^{+3.3}$   & $0.91_{-0.01}^{+0.01}$ & $0.26\pm0.02$ &  \\
$H2$ & $57.3_{-4.3}^{+4.1}$ & $52.1^{+2.6}_{-2.3}$ & $145.4_{-3.4}^{+2.8}$  & $0.89_{-0.02}^{+0.02}$ & $0.29\pm0.01$  & 2162.9/2138  \\
$L2$ & $26.5_{-1.0}^{+1.8}$ & $100^{+*}_{-47.6}$ & $122.2_{-4.0}^{+5.3}$ & $0.89_{-0.02}^{+0.01}$ & $0.22\pm0.02$ & \\
\hline
$L1$ & $18.2_{-0.4}^{+0.5}$ &                     &                        & $0.89_{-0.01}^{+0.01}$ & $0.25\pm0.01$ &   \\
$H1$ & $32.5_{-3.9}^{+14.1}$ &                     &                        & $0.91_{-0.01}^{+0.01}$ & $0.26\pm0.01$ &  \\
$H2$ & $29.7_{-5.8}^{+4.7}$ & $88.1^{+*}_{-22.4}$& $137.3_{-2.7}^{+3.9}$  & $0.89_{-0.01}^{+0.01}$ & $0.28\pm0.01$  & 2184.1/2148  \\
$L2$ & $18.0_{-0.6}^{+0.7}$ &                     &                        & $0.88_{-0.01}^{+0.01}$ & $0.24\pm0.01$ & \\
\hline
$L1$ &        & $46.6^{+9.2}_{-5.3}$ &                      & $0.88_{-0.01}^{+0.01}$ & $0.23\pm0.01$ &  \\
$H1$ &        & $99.9^{+*}_{-32.7}  $ &                      & $0.91_{-0.01}^{+0.01}$ & $0.27\pm0.01$ &  \\
$H2$ & $25.6_{-2.9}^{+4.8}$ & $90.1^{+*}_{-25.5}$& $136.2_{-5.6}^{+3.8}$  & $0.89_{-0.01}^{+0.01}$ & $0.29\pm0.01$ & 2171.0/2148  \\
$L2$ &        & $47.5^{+8.0}_{-5.7}$   &                      & $0.87_{-0.01}^{+0.02}$ & $0.22\pm0.01$ & \\
\hline
$L1$ &        &  & $126.2_{-4.6}^{+6.8}$  & $0.88_{-0.01}^{+0.01}$ & $0.23\pm0.01$ & \\
$H1$ &        &  & $138.5_{-4.6}^{+6.1}$  & $0.90_{-0.01}^{+0.01}$ & $0.27\pm0.01$ &   \\
$H2$ & $24.0_{-1.9}^{+5.7}$ & $86.7^{+*}_{-33.0}$& $139.3_{-4.4}^{+6.3}$  & $0.89_{-0.01}^{+0.01}$ & $0.29\pm0.01$  & 2171.8/2148  \\
$L2$ &        &  & $124.2_{-4.6}^{+6.3}$  & $0.87_{-0.01}^{+0.02}$ & $0.22\pm0.01$ &    \\
\hline
$L1$ &        & $ $ &                      & $0.90_{-0.01}^{+0.01}$ & $0.22\pm0.01$ &  \\
$H1$ &        & $ $ &                      & $0.90_{-0.01}^{+0.01}$ & $0.26\pm0.01$ &  \\
$H2$ & $28.3_{-4.6}^{+*}$ & $67.2^{+*}_{-18.0}$& $136.8_{-5.3}^{+6.0}$  & $0.89_{-0.01}^{+0.01}$ & $0.29\pm0.01$ &  2206.1/2151  \\
$L2$ &        & $ $ &                      & $0.88_{-0.01}^{+0.02}$ & $0.21\pm0.01$ &  \\
\hline
$L1$ &        & $ $ &                      &  & $0.22\pm0.01$ &  \\
$H1$ &        & $ $ &                      &  & $0.26\pm0.01$ &   \\
$H2$ & $27.5_{-4.3}^{+*}$ & $67.2^{+*}_{-18.3}$& $137.1_{-6.0}^{+5.5}$  & $0.89_{-0.01}^{+0.01}$ & $0.29\pm0.01$ & 2229.3/2154  \\
$L2$ &        & $ $ &                      &  & $0.21\pm0.01$ &   \\
\hline
\end{tabular}
\begin{tablenotes}
\item[] The columns are: (1)-(5) the coronal radius r$_{cor}$ (in unit R$_{g}$=GM/c$^2$),
the optical depth $\tau$ of the soft Comptonization component, the electron temperature $kT$ for the soft Comptonization component (soft excess) in eV,
the fraction of the power below r$_{cor}$ that is emitted in the hard Comptonization component, and the Eddington ratio of the {\it optxagnf} model;
(6) the $\chi^2$ and degree of freedom.
\end{tablenotes}
\end{threeparttable}
\end{table}

\end{document}